# Dynamic Feature Description in Human Action Recognition

Ruoyun Gao

s0733369

Computer Science

Supervisors: Michael S. Lew
                Ling Shao (Philips)

# Acknowledgement

I wish to thank all those who helped me. Without them, I could not have completed this project.

Firstly, I wish to express my deepest gratitude to Dr. Michael S. Lew for his valuable suggestions and unwavering support during my research project and graduation project. Without his encouragement and trust, I could not have finished my project so smoothly.

I would also specially thank my company supervisor Dr. Ling Shao who gave me the great opportunity to work in Video Processing & Analysis, Philips Research and provided insightful guidance and continuous encouragement during my project.

Many thanks to all my colleagues who helped, supported and accompanied me to conquer all the difficulties in my project.

# Table of Contents

# List of Figures

# Abstract

This thesis aims to present novel description methods for human action recognition. Generally, a video sequence can be represented as a collection of spatial temporal words by detecting space-time interest points and describing the unique features around the detected points (Bag of Words representation). Interest points as well as the cuboids around them are considered informative for feature description in terms of both the structural distribution of interest points and the information content inside the cuboids. Our proposed description approaches are based on this idea and making the feature descriptors more discriminative.

In this thesis, we propose and validate three types of description methods: the first is *projected 3D Shape Context (projected 3DSC)* which is derived from the original shape context and makes use of the structural distribution of interest points; the second are the *Transform based description methods* which are widely exploited in image processing and utilizes the appearance information of images; the third is *Correlogram of Oriented Gradient* (*COG*) which is built from the spatial temporal gradients of each cuboid taking advantage of both the spatial structure and appearance information.

The proposed methods are tested in the very challenging and well studied KTH dataset. The *projected 3DSC* improves the classification accuracy by 10% compared to that of the original 3DSC. And the *Wavelet Transform based descriptor* achieves as high as 93.89% recognition rate, which is better than any of the state-of-the-art methods. *Correlogram of Oriented Gradient* achieves 15% better than the popular Histogram of Oriented Gradient (HOG). Further more, we validate the efficiency of the Wavelet Transform based method on a more realistic and challenging human action dataset: the Hollywood dataset.

# 1 Introduction

Nowadays, more and more people record their daily activities using digital cameras, and this brings the enrichment of video sources on the internet, and also causes the problems of how to categorize the existing video sources and how to classify new input videos according to their action classes. Apparently, categorizing these videos for future processing is a time-consuming task if it is done manually, and recognizing certain actions from scenes of interest in real movies is impossible to accomplish by manual effort. That is why these problems attract a lot of researchers' attention. They attempted to build a pattern recognition system, which trains on the feature descriptors extracted from the training videos, and enables the computer to identify the actions of new videos automatically. The objective of this thesis is to optimize the feature description part of the widely used recognition scheme.

## 1.1 Motivation

The development of computer vision has encouraged the occurrence of different novel recognition methods in both 2D images and 3D video sequences. Although it is still challenging to recognize a specific object from a dataset of images due to viewpoint change, illumination, partial occlusions, and intra-class difference and so forth, yet many successful methods have been proposed [3,39,54]. But for the video recognition problem, the current methods still need improvement, especially for the realistic movies which have more individual variations of people's posture and clothes, dynamic background, and partial occlusion, etc. Intuitively, a straightforward way is comparing an unknown video with the training samples by computing correlation between the whole videos [2]. This approach makes good use of geometrical consistency, but it is not feasible in dealing with camera motions, zooming, intra-class difference and non-stationary background, etc.

To conquer these deficiencies, a lot of researchers focus on the part-based approaches for which only the 'interesting' parts of the video are analyzed other than the whole video. These mentioned 'parts' can be trajectories or flow vectors of corners, profiles generated from silhouettes and spatial temporal interest points. We adopt the recognition scheme that utilizes spatial temporal interest points, which are robust to viewpoint variation, moving background, zooming and object deformation [3]. However, the recognition rate is still constrained due to the inefficient and unreliable description methods. Our contributions are to account for these issues by the proposal of novel and effective description methods. The proposed methods are:

1. **Projected 3DSC**: This description method makes good use of structural distribution of the spatial temporal interest points, and it produces very precise and short descriptors. Its capacity of representing the discrimination and commonness among actions makes it superior to the original 3DSC.

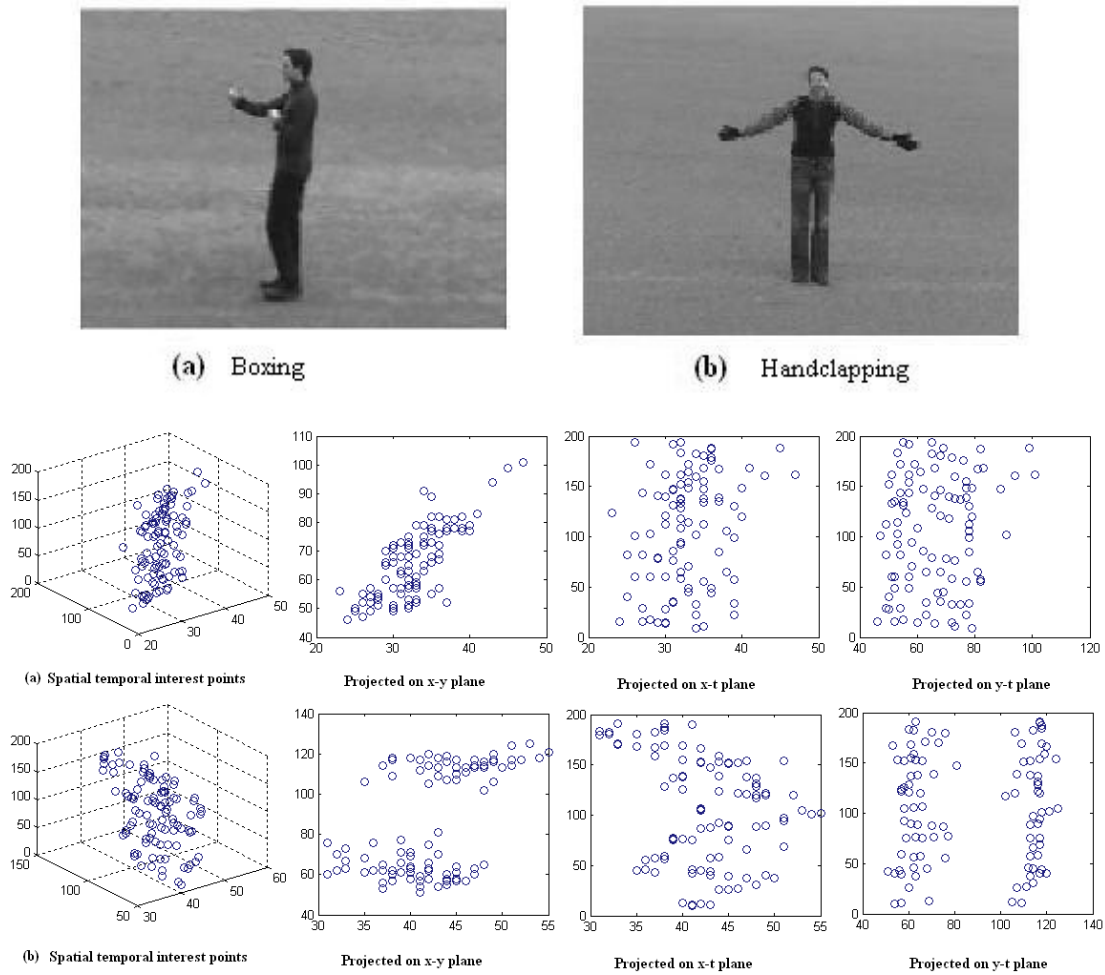

Fig. 1  Subject (a) is performing 'boxing', while subject (b) is performing 'handclapping'. Their spatial temporal interest point distributions (point clusters) show no clear feature differences. However, after we project these interest points to x-y plane, y-t plane and x-t plane respectively, the differences are obviously displayed.

2. **Transform based descriptors**: We introduce the transform techniques to the action recognition area, and they continue their good performance in signal processing area, especially the Discrete Wavelet Transform, which generates the shortest descriptor length but is extremely compact and informative, and therefore outperforms any of the state-of-the-art algorithms.

3. **COG**: The Correlogram of Oriented Gradient (COG) based on the 3D gradients is an enhanced version of the popular Histogram of Oriented Gradient (HOG) by incorporating the spatial co-occurrence of gradients. Therefore COG is a straightforward way of exploring both appearance and structural information.

According to the recognition scheme, we build a model out of a collection of feature descriptors, and this model is an input to the further classification and recognition phase. For classification, numerous discriminative classifiers have been well exploited, such as Support Vector Machine (SVM), K Nearest Neighbor (KNN), and generative models such as pLSA [4] and LDA.

## 1.2 Approaches

### 1.2.1 Shape context

It is a common sense that each human action has a similar style of moving arms, legs or other body parts. For instance, when one man is running, it is natural to swing his arms, and everyone swings his arms in a similar style. This property contributes a possibility of modeling each action by the general shape of all the moving parts. Based on this perspective, we use the shape context description method to represent each action's 'shape' structure. Firstly we extract certain number of spatial temporal interest points that have the strongest motion from each sequence, and these interest points constitute a point cluster that can be treated as a discriminative representation of one certain sequence. Secondly, we apply our modified shape context on these point clusters and build a model from all training sequences. Finally in the matching phase, each test sequence is compared with the action model and gives the correct action category it belongs to.

The primary contribution in this part of work is the proposal of a novel but efficient approach making use of the structural information of interest points. The distribution of the extracted points (point cluster) indicates the representative features of one certain action, and shapes of interest points are expected to be similar between the same actions and distinguished between any two different actions. However, the 3D point cluster is actually not discriminative enough to differentiate two actions. In Fig. 1, the 3D point clusters of *'Boxing'* and *'Handclapping'* exhibit no obvious distinctions. In order to clearly interpret the inter-action ambiguity, we propose to project the point cluster to certain numbers of lower dimensional planes. This is to achieve more representative motion features from different viewpoints. For simplicity, we choose 3 orthogonal projection planes, which are *x-y*, *y-t*, and *x-t* planes respectively, in our application; and the detected 3D point cluster is projected to each of these 3 planes. As Fig. 1 shows, the shapes of each action after projection display significant and clear differences between two actions. Therefore, we translate the projected shape information to description vectors by making use of the structural relation among these points. We consider each projection plane a 2D image and apply the original 2D Shape Context on it. The original shape context algorithm [35] is to detect the edge points of the objects and then for each edge point, a shape context descriptor is computed on the remaining points. However, we apply shape context on the detected interest points other than the edge points. Similarly, the shape context of each point is flattened to be a description vector. We call our method the '*projected 3DSC*'. This method highlights the information along each projection plane and provides more discrimination during the model building and matching phases.

### 1.2.2 Transform based descriptors

The aim of this work is to bring up new and discriminative feature description methods to make features more reliable in representing sequences and further improve the recognition rate. Based on the wide usage and good performance of time frequency transformation techniques in the image processing area, we introduce several transformation methods into the human action recognition application. In our implementation, we utilize the Discrete Fourier Transform (DFT), Discrete Cosine Transform (DCT) as well as the Discrete Wavelet Transform (DWT) due to their good performance in image and video processing.

A sequence is represented as features aggregated from a collection of spatial temporal interest points. And at each detected interest point, a cuboid is extracted which contains the spatial-temporally windowed pixel values. These cuboids, containing appearance information of video events, are the raw materials for further feature description. In our application, we select three orthogonal planes intersected in the center of each cuboid, and apply transformation techniques on each plane. The resulted coefficients after transformation are compact and discriminative, thus, all the coefficients from every cuboid are concatenated to constitute a descriptor for this cuboid. Although two subjects performing the same action may vary significantly in terms of overall appearance and motion, yet the extracted interest points manifest the uniqueness of the same action. Under this assumption, the obtained descriptors from cuboids can be clustered to limited number of video words. These video words together form a *codebook* which is used to build the learning model, and other feature descriptors are assigned to certain members of the codebook. In this way, it simplifies the usage of all the extracted features while still maintains their richness in information [6]. This way of representation is known as the 'Bag of Words'. And it ignores the geometric arrangement between visual features and an action video is represented as a histogram of the number of occurrences of particular video words [4].

### 1.2.3 Correlogram of Oriented Gradient

We borrow the color histogram idea from image retrieval [41, 40] and apply it on the spatial temporal gradients (3D gradients) in video. In our implementation, the brightness gradients are calculated within each cuboid along x, y, and t, and the 3 gradients are computed as the ratios of 2 gradients against another one. In this way, the number of parameters is reduced to be two. For each cuboid, we build a histogram from the combination of the two ratio sets (HOG). Generally, histogram is the occurrence of certain features in histogram bins, thus it gives a coarse distribution of the features. However this definition displays the inherent deficiency of histogram, because it only has the global feature distribution without local positional information of features. Fig. 2 gives an example of this issue using color as feature. The traffic icon and 'Bagua' are considered the same by color distribution (equal amount of 'red' and 'white'). Actually, positions of 'red' and 'white' are completely inconsistent and the shapes are different as well. Inevitably, the Histogram of Oriented Gradient (HOG) also has this shortcoming. To add positional information of gradients, we propose to adopt correlogram on the gradients, which is to capture the spatial co-occurrence of a pair of the combined gradient ratio sets with regard to distance (COG). Therefore it incorporates the appearance information (gradients) as well as the spatial correlations (positions of gradients) and is robust to small geometric deformation.

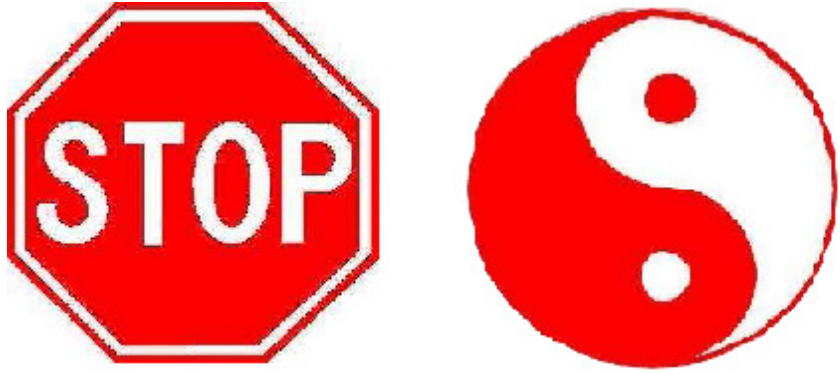

Fig. 2  Example of same color distribution but different shapes.

## 1.3 Organization of this thesis

In this paper, we propose three kinds of description methods and experimentally validate their efficiency under different testing conditions. Section 2 reviews the research background and the state-of-the-art methods in human action recognition. We illustrate the framework of human action recognition algorithm in Section 3. And the shape context related approaches are introduced in Section 4. In Section 5, we apply the transform domain techniques on action recognition and evaluate their performance. Section 6 compares the efficiency of HOG and COG. To further testify our proposed methods, we present the detailed experiments achieved on the realistic Hollywood dataset in Section 7. Finally, we conclude our work and point out possible directions for improvement in Section 8.

# 2 Literature study

Piles of work have been done in human action recognition and localization to save manual work and accelerate processing efficiency. Although there is no perfect algorithm to deal with all the problems occurring in action recognition, such as viewpoint change, complex background activities and partial occlusion, yet the current work has shown quite promising results under different scenarios. One branch of work is to utilize the global information of the subject subtracted from video data, for example the correlating optical flow measurements from low resolution videos proposed by *Efros et al.* [19] which segment and stabilize each human figure and annotate each action in the resulted spatial temporal volume. Another part of the work attempts to track the body parts and use the motion trajectories to discriminate different actions [20, 21]. In their implementations, certain feature points are located in a frame-by-frame manner, and the tracks of these points show many discriminative properties, such as position, velocities and appearance. However, the methods mentioned above are sensitive to partial occlusion and use much redundant information that is computationally expensive. The drawbacks of these methods accelerate the prosperity of the part-based approaches, especially the space-time interest point detectors proposed by *Laptev et al.* [5] and *Dollár et al.* [6]. *Laptev et al.* [5] propose a space-time interest point detector based on the idea of the *Harris* and *Förstner* interest point operators. Their approach detects local structures in space-time where the image values have significant local variations in both space and time. *Dollár et al.* [6] use a set of separable linear filters detecting interest points with strong motion. This method is designed to respond to complex motion of local regions and the space-time corners. It detects more number of interest points than *Laptev*'s approach, which makes it more reliable in processing videos with limited frames. *Ke et al.* [7] apply spatial temporal volumetric features that efficiently scan video sequences in space and time and detect interest points over the motion vectors.

The detected spatial temporal interest points can be used to construct 3D shape context descriptors based on the concept that one certain action provides similar structural distribution of interest points [38]. This structural information refers to the configuration of entire shape with regard to a reference point [35]. On the other hand, *Dollár et al.* [6] extract spatial temporal video patches (cuboids) around each interest point, which contain the appearance information of video event, and can be further processed to provide unique features. This appearance information can be brightness gradient, optical flow and graphical shape etc.

However, we discovered that more compact and informative description approaches can be applied in the human action recognition area. Therefore, in the following sections the main objective is to present and validate the projected 3DSC in action recognition problem; and also to introduce transform domain techniques into this field, due to their great performance in signal processing; and finally to prove correlogram outperforms histogram in representing features.

## 2.1 Shape context

The definition behind shape context is mainly referring to two aspects as elaborated in [30] and [31]. The first aspect is feature based approach, which takes advantage of the spatial arrangements of the extracted features, for instance the silhouette elements or junctions. The second aspect is brightness based approach which uses pixel values directly. Many years of work have been done based on the first method mentioned above. *Sharvit et al.* [32] attempt to capture the part structure of the shape in the graph structure of the skeleton. *Gdalyahu et al.* [33] use 1D silhouette curves for matching between two objects. However, silhouette based methods have intrinsic drawbacks due to the fact that silhouette ignores the internal contour and is very hard to extract from complex background. One alternative is to consider the object shape as a set of points and use edge detector to extract points as explained in [34]. *Belongie et al.* [35] propose to capture a subset of edge points and for each point build a shape context which is a histogram of the relative coordinates of the remaining points in shape matching and object recognition. Derived from the idea of [35], *Kortgen et al.* [36] extend 2D Shape Context into 3 dimensional object matching. *Shao and Du* [37] further extend *Belongie et al.*'s idea into 3D video sequences, which is to build descriptors containing the spatial temporal distribution of the remaining interest points regarding to a reference point. This approach, named as 3 Dimensional Shape Context (3DSC), performs well in action recognition with scale and translation invariance.

## 2.2 Transform based descriptors

Transform domain techniques have been widely used in the image processing field, such as image compression, enhancement and segmentation etc. There are three well known transforms in this area, which are Discrete Fourier Transform (DFT), Discrete Cosine Transform (DCT) and Discrete Wavelet Transform (DWT). Fourier Transform is to analyze a signal in the time domain for its frequency content and translate it into a function in the frequency domain. The DFT estimates the Fourier transform of a function from a finite number of its sampled points [24]. DCT is similar to DFT, but only using real data with even symmetry. DWT decomposes a discrete signal into a set of discrete basis functions (wavelets) instead of sine and cosine waves used in Fourier Transform, and it captures both the frequency information and the space information.

Recently, DCT and DWT have been widely exploited by modern image and video coding standards, such as JPEG, MPEG etc. [9, 15]. *Smith et al.* [50] propose a texture classification method based on the variance and the mean absolute values of the DCT coefficients calculated over the entire image. *Climer et al.* [51] propose a quadtree-structure-based method using the DCT coefficients on the nodes of the quadtree as image features. On the other hand, Wavelets have been used in detecting salient points representing local properties of images in content-based image retrieval, which is proposed by *Sebe et al.* [1]. *Schneiderman et al*. [52] describe a statistical method for 3D object detection by histogramming the subset of Wavelet coefficients and their position on the object. *Gonzalez-Audicana et al.* [53] use the multiresolution Wavelet decomposition to execute the spatial detail information extraction for image fusion.

The primary advantage of these transformation methods is the removal of redundancy between neighboring pixels [19], and it leads to uncorrelated transform coefficients which can be encoded independently. Among all the transform techniques, Wavelet Transform has the best performance by localizing in frequency and also in space. Therefore Wavelet Transforms are desirable to deal with signals which are localized in time or space (speech or imagery) [8, 22].

## 2.3 Correlogram of Oriented Gradient

Histogram is easy to compute and invariant to rotation and translation of the image content due to its capture of global features. Particularly, Histogramming the gradient information of images has been widely used in the past decades in image retrieval and indexing [42, 43, 44]. The principal idea behind these algorithms is that appearance and shape of local objects can be characterized well by the distribution of local intensity of gradients, even without the position knowledge of each gradient. Lowe's *Scale Invariant Feature Transformation (SIFT)* [42] is a milestone of making use of gradient information. It is to extract and match scale invariant keypoints by building sub-region descriptors and adopting the local spatial gradient histogramming and normalization.

Correlogram has been long used in computer vision and image retrieval areas. *Julesz* [46] uses gray level spatial dependence and *Haralick* [47] proposes to describe two-dimensional spatial dependence of gray values by a multi-dimensional co-occurrence matrix. In [48], *Huang et al.* refine the co-occurrence matrix by maintaining only the correlation of pairs of color values as a function of distance. This statistic matrix captures the relationship between the spatial correlations of all possible pairs of features as a function of distance, while histogram only captures the global distribution of features from images. *Savarese et al.* [54] propose to use the correlogram for capturing the spatial arrangement of pixel labels (visual words) as a representation of an image. Again, *Savarese et al.* [49] extend correlogram to the spatial temporal domain.

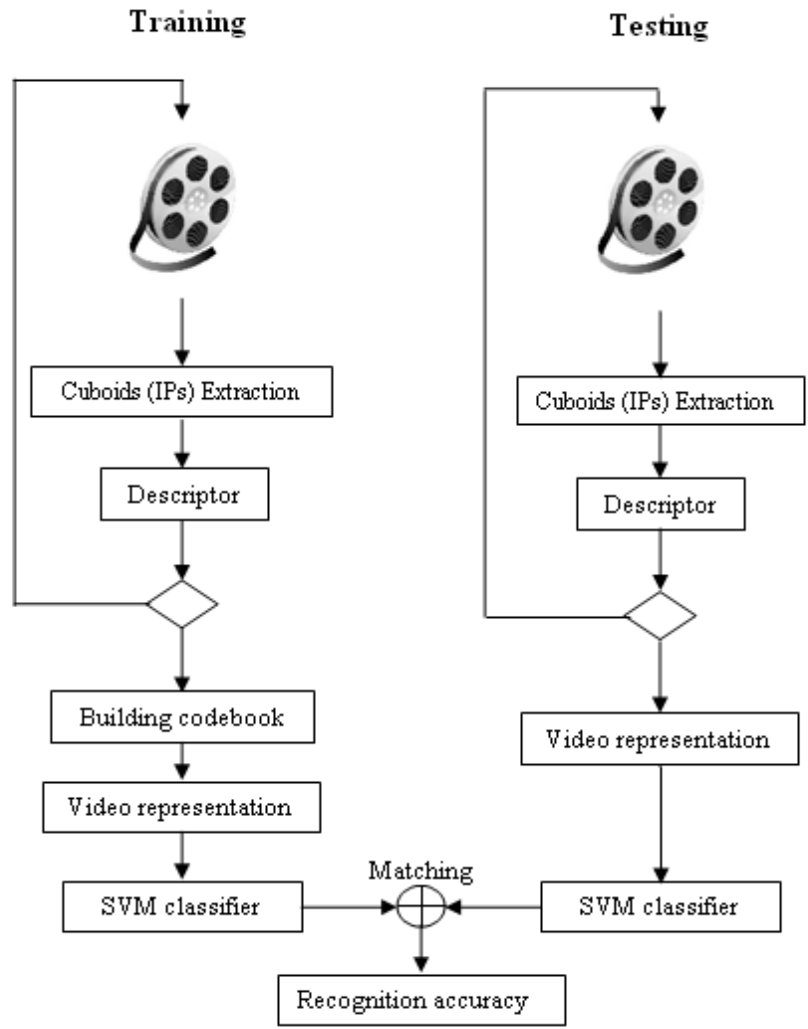

Fig. 3  Flowchart of the algorithm. IP is short for Interest Points.

# 3 Human action recognition framework

The whole recognition process can be divided into two phases: the training phase and the testing phase. During the training phase, as the flow chart shown in Fig. 3, the interest points as well as the cuboids surrounding them are extracted by some interest point detector from the training sequences, and then descriptor of each sequence is generated by the structural distribution of interest points or the appearance information embedded in each cuboid. Descriptors from all training sequences are gathered together for further clustering by K-means which uses Euclidean distance as the clustering metric. The cluster centers are represented as the video words and they constitute the codebook. Each feature descriptor is assigned to a unique video word based on the distance between the descriptor and cluster centers. And the codebook membership of each feature descriptor is utilized to create a model representing the characteristics of each class of the training sequences.

During the testing phase, we follow the same steps to extract interest points, build descriptors and assign codebook membership as those done during the training phase. Then Support Vector Machine (SVM) and K Nearest Neighbor (KNN) classifiers are adopted to classify each testing sequence to the most probable action type according to the model built in the training phase, and the correctly classified sequences against all sequences give the finial recognition accuracy.

## 3.1 Interest points detection

Interest points from a video sequence are localized not only along the spatial dimensions *x* and *y* but also the temporal dimension *t*. Currently there are three types of detection approaches: static features based on edges and limb shapes, dynamic features based on optical flow measurements and spatio-temporal features obtained from local video patches. *Dollár et al.*'s method [6] is an instance of the third type. It uses separable linear filters and generally produces a high number of points. It assumes a stationary camera or a process that can account for camera motion. The response function is defined as follows:

$$R = (I*g*h_{ev})^2 + (I*g*h_{od})^2 \quad (1)$$

where *g(x, y; σ)* is the 2D Gaussian smoothing kernel, applied only along the spatial dimensions, and $h_{ev}$ and $h_{od}$ are a quadrature pair of 1D Gabor filters applied temporally. These are defined as $h_{ev}(t;\tau,\omega) = -\cos(2\pi t\omega)e^{-t^2/\tau^2}$ and $h_{od}(t;\tau,\omega) = -\sin(2\pi t\omega)e^{-t^2/\tau^2}$. In all cases the authors suggest $\omega = 4/\tau$, effectively making the response function R only two parameters *σ* and *τ*, corresponding roughly to the spatial and temporal scales of the detector. The response function gives strong response to periodic motion as well as the spatio-temporal corners. In general, any region with spatially distinguishing characteristics undergoing a complex motion can induce a strong response. Areas undergoing pure translational motion or without spatially distinguishing features will in general not induce a response, as a moving smoothed edge will only cause a gradual change in intensity at a given spatial location. The space-time interest points are extracted

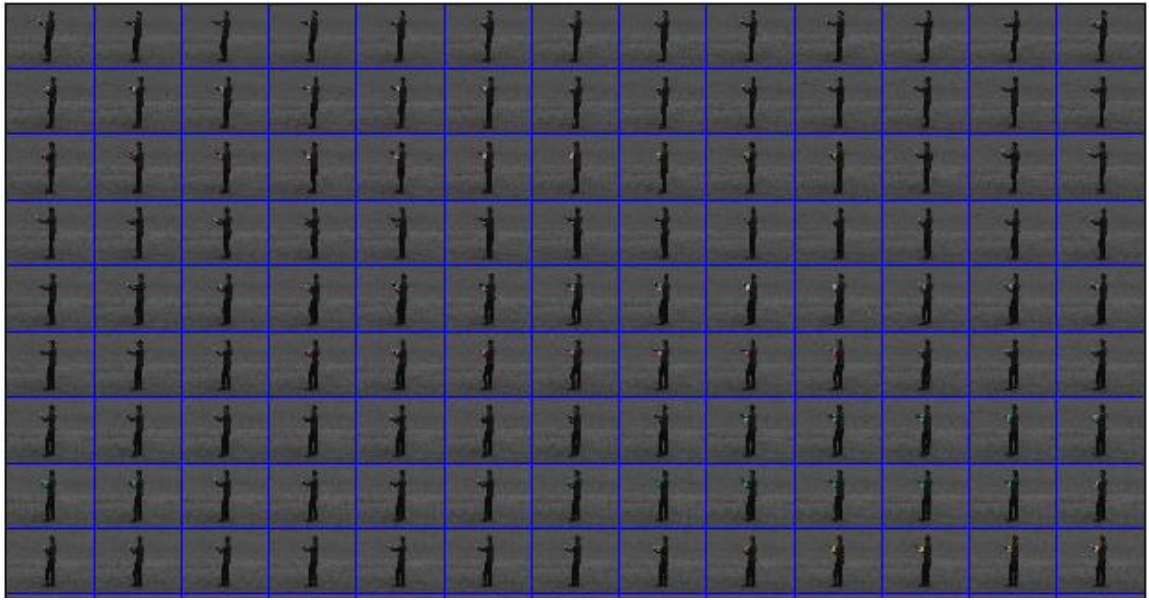

(a). Illustration of the location of the extracted spatial temporal interest points for action *'boxing'*. The highlighted points indicate the interest points of each frame.

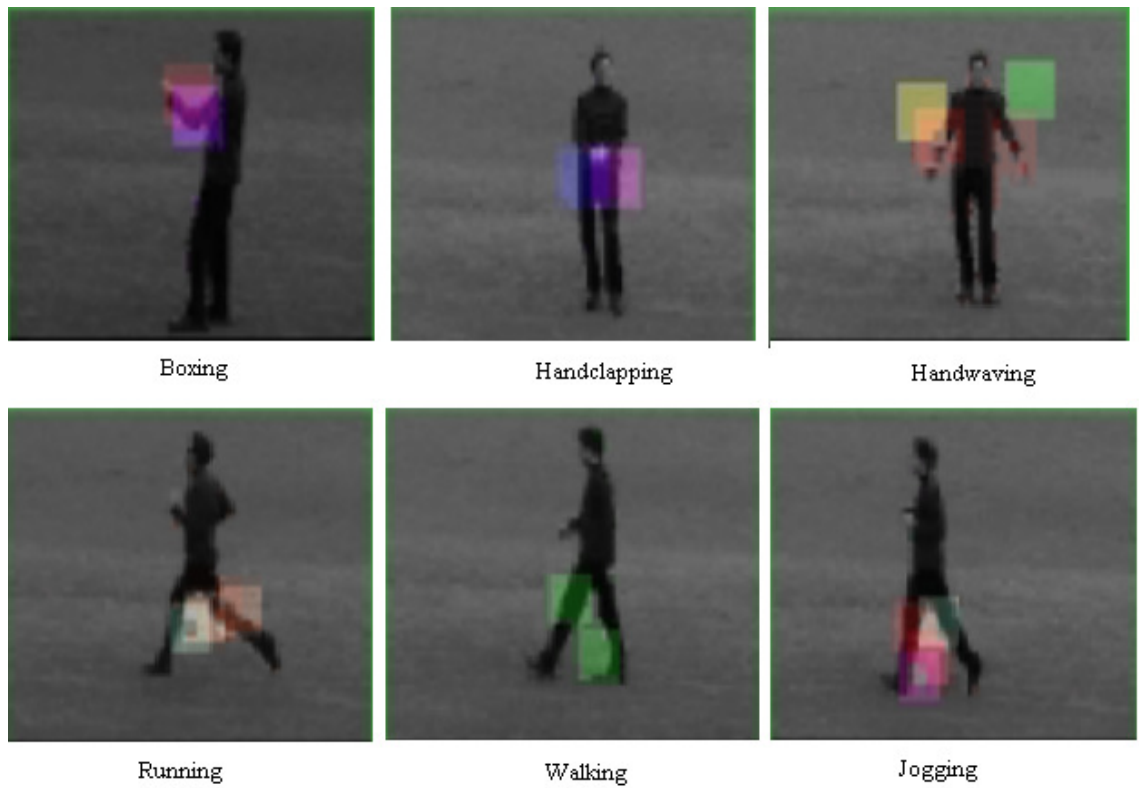

(b). Cuboids generated around interest points for each action.

Fig. 4  Illustration of interest points and cuboids.

around the local maxima of the response function and could be regarded as low-level action features. Fig. 4 (a) shows the detected interest points in a *'Boxing'* sequence. The highlighted points correspond to local maxima of response function and are considered as spatial temporal interest points. Fig. 4 (b) shows the cuboids extracted from each action type displayed in one frame, and it also proves actions can be discriminated by these cuboids due to the fact that different actions generate different cuboids distribution.

As mentioned in [6], a cuboid (spatial temporal video patch) is extracted around each interest point and it contains spatio-temporally windowed pixel values (Refer to Fig. 5). The size of cuboid is set to be of approximately six times the scale at which they were detected. The information contained in each cuboid is utilized to form a representative descriptor and moreover to build the action training model. The locality of cuboids facilitates the feature extraction which means preprocessing steps are not needed, such as the foreground subtraction and figure tracking and alignment etc. Relying on each individual cuboid, we can obtain the appearance information from cuboid itself as well as structural information from the distribution of all cuboids (interest points).

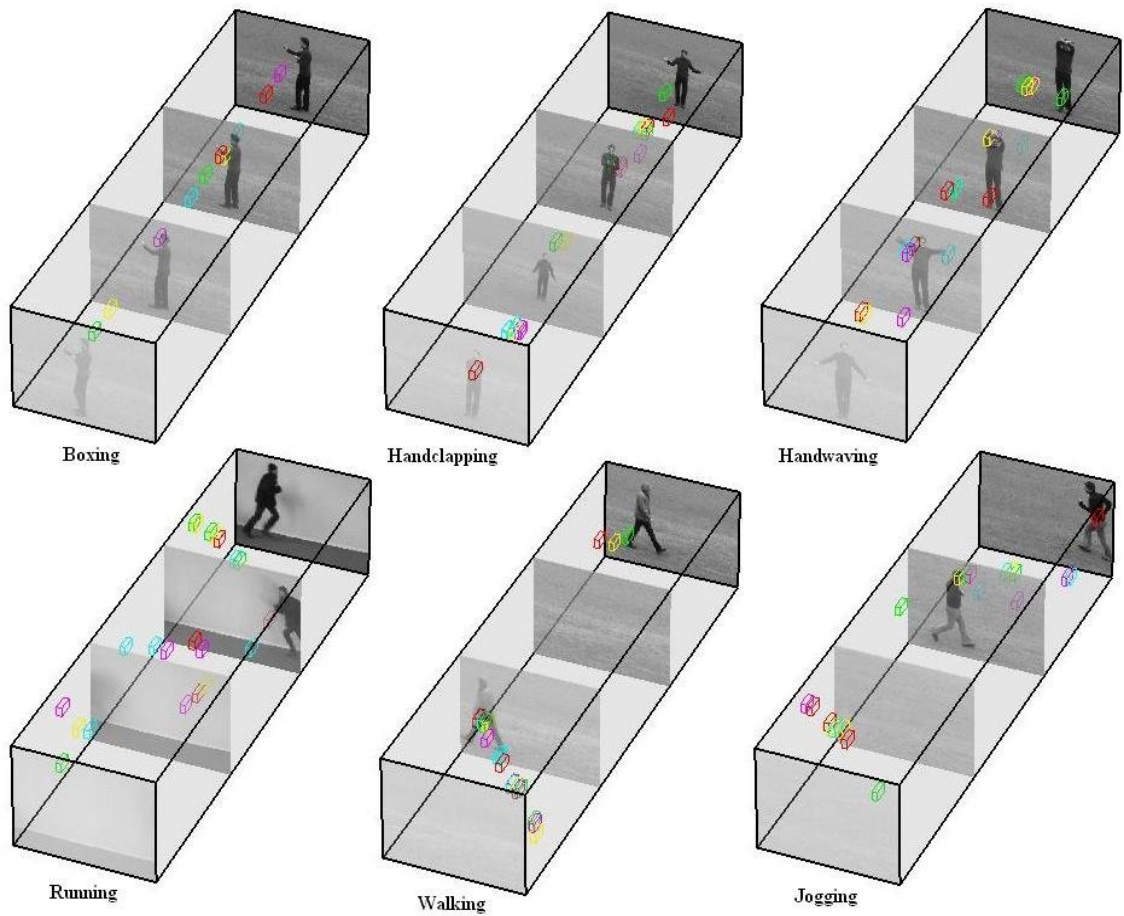

Fig. 5  Stereo views of cuboids extracted around each spatial temporal interest point for each action. The cylinder indicates an action sequence, and each shows 4 frames for simplicity. Colored cuboids are displayed to express the idea that different actions have different interest points (cuboids) distribution.

## 3.2 Feature description methods

### 3.2.1 Shape context

Shape context has proven its efficiency in shape matching and object recognition by solving for correspondences between points on two shapes and using the correspondences to estimate an aligning transform. A lot of approaches derived from the original two-dimensional shape context are currently in use in many areas and obtain good reputation for extracting the structural features efficiently.

**(1) 3DSC**

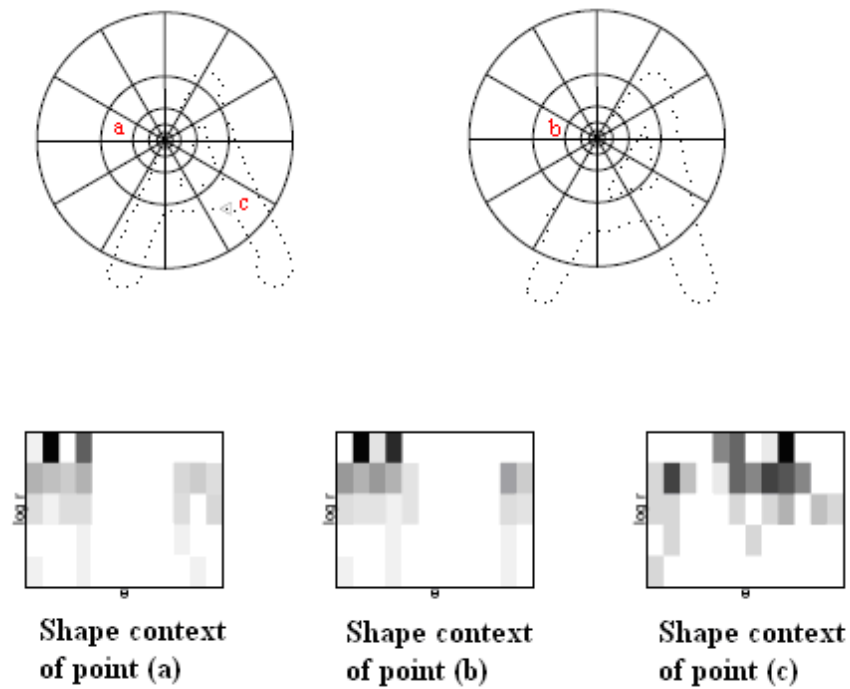

Fig. 6  Diagram of log-polar histogram bins for reference points and their shape context histograms.

3D Shape Context (3DSC) is an extension of 2D Shape Context, and captures the occurrence of spatial temporal points with regard to a reference point. The original shape context applied in images is proposed by *Belongie et al.* [35]. Their method is to use log-polar histogram bins to compute the shape contexts for each point, and shape context is a coarse histogram of points' occurrence in each bin, shown in Fig. 6. The darker the color in a shape context means more points are located in this bin. Apparently, shape contexts of point (a) and (b) have quite similar bin color distribution and are considered as corresponding. While point (c) has a quite different shape context with either point (a) or (b), thus point (c) is not corresponding to (a) or (b). For more details refer to [35].

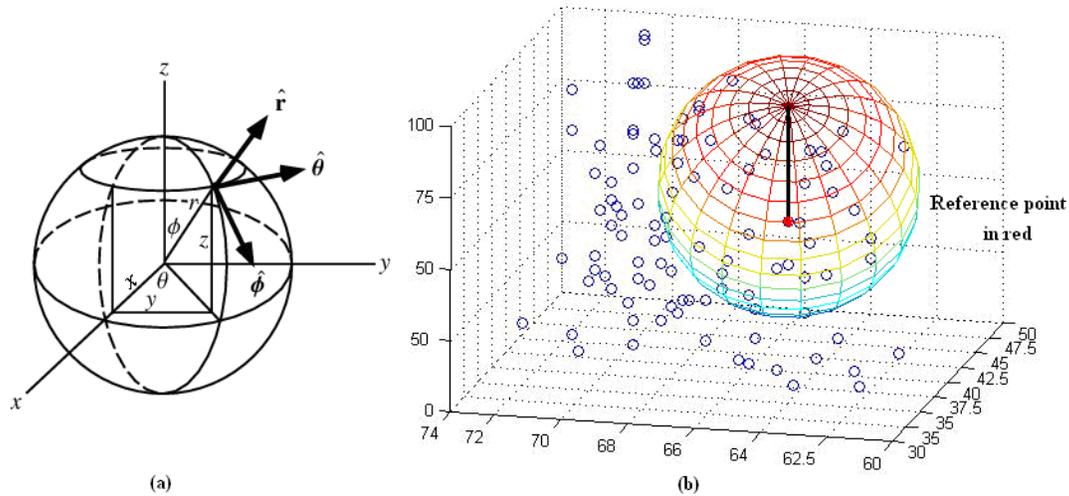

Fig. 7  (a) is the sphere kernel S ($\varphi$, $\theta$, r). (b) shows interest point cluster in the space and time, and a sphere kernel centered on one reference point.

In [37], *Shao and Du* extend the circle kernel to a sphere kernel $S(\varphi, \theta, r)$ (Fig. 7 (a)) partitioned into several cubes by angular parameter $\varphi$ and $\theta$ and radial parameter $r$. For a sequence, the interest points are detected from space as well as time. Thus we center the sphere kernel on each interest point and generate the histogram of occurrence of interest points within each cube, as illustrated in Fig. 7 (b). In our implementation, each description vector of an interest point is a concatenation of each cube in the sphere kernel.

**(2) Projected 3DSC**

The 3DSC uses the actual position of each interest point to form its shape context. Therefore even small variation of positions may cause points falling into their neighboring cubes, and accordingly generate a different shape context descriptor. Since each subject has his own habit of performing certain actions, for instance how much to swing his arms when running and how big his step is, therefore it is obvious to notice that the positions of interest points extracted from these body parts vary among different subjects even when they are doing the same action. In this case, 3DSC is too sensitive to these common variations and may fail to give a reliable representation of one sequence and finally misclassify one action to others. To loose this position constraint, we project these 3-dimensional points to three orthogonal planes by eliminating *x*, *y*, or *t* dimension respectively. Therefore, small position variations along *x*, *y*, *t* dimensions are decreased, and the tolerance for these inevitable difference between subjects is increased. As illustrated in Fig. 8, the space time interest points as well as their projected point distributions are displayed for six actions. Clearly, the projected point distributions manifest the main characters of each action and can generate discriminative shape context descriptors. Compared among these 6 actions, the projected point distributions also give the commonness between similar actions, for instance, *'Running'*, *'Walking'* and *'Jogging'* have similar points distribution when projected to *x-t* plane due to the fact that these three actions are quite similar.

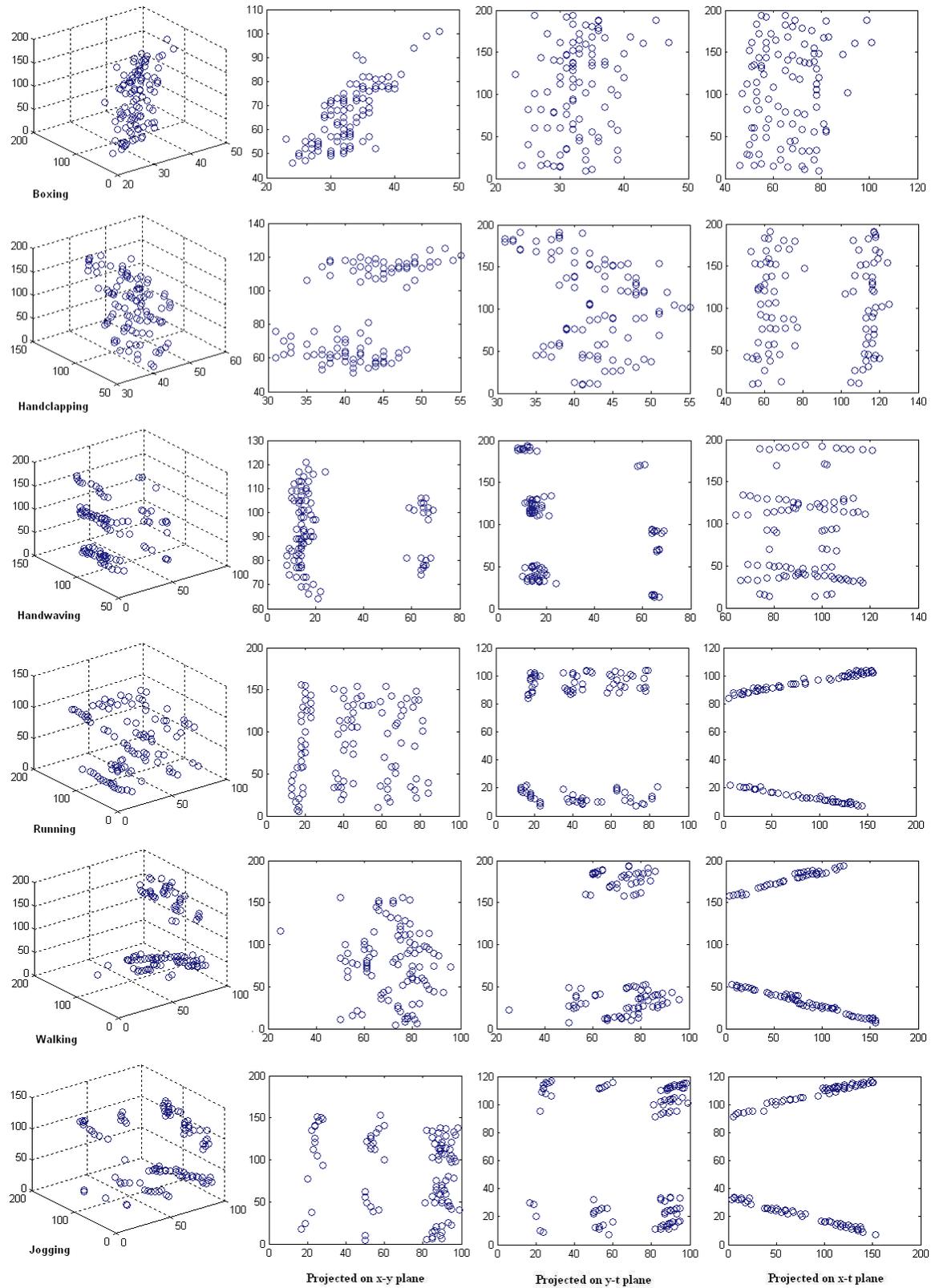

Fig. 8  Spatial temporal interest points' distributions and the projection results for each action.

In conclusion, projected 3DSC is superior in highlighting the discrimination and preserving the commonness between any two actions. The discrimination facilitates the application of the original 2-dimensional shape context and guarantees the recognition accuracy. And the commonness shows our algorithm's reliability in characterizing actions. In our implementation, we apply the original 2D Shape Context on each projection plane to construct sub-descriptor of each plane and then concatenate the sub-descriptors to form the final descriptor for this sequence.

### 3.2.2 3D gradient

The 3D gradient description method is proposed by *Dollár et al.* [6]. They smooth the cuboids at different scales and then calculate the brightness gradient at each location ($x, y, t$). The 3D gradient descriptor consists of three channels ($G_x, G_y, G_t$) and each has the same size as that of the cuboid. In their implementation, the brightness gradients are concatenated to be a vector representing one cuboid. However, this concatenation makes the descriptor potentially sensitive to small cuboid perturbation, and a dimensionality reduction technique is required to accommodate the memory limit of computers. In this thesis, we utilize Principal Component Analysis (PCA) as the dimensionality reduction approach, and the reduced dimension is fixed to be 100.

### 3.2.3 Transform based descriptors

$$X \xrightarrow{T} \tilde{X} \xrightarrow{process} \tilde{Y} \xrightarrow{T^{-1}} Y$$

Fig. 9  General signal transformation process.

Classic transformation methods have been widely and efficiently used in the image processing area, such as image de-noising, image segmentation, feature detection and compression. Fig. 9 gives the general process of signal transformation. *X* is any signal, and *T* is some transform technique and *'process'* in the middle is usually non-linear. $T^{-1}$ is the reverse transformation. Simply speaking, transformation techniques are to convert a signal to different frequency components. There are many well explored transformation techniques in the signal processing field, such as the Fourier Transform and the Wavelet Transform.

Fig. 10 illustrates the structure of one single cuboid. For each cuboid, we select three orthogonal planes intersecting in the middle of cuboid and apply different transformation techniques on these planes. The transformed coefficients are the discriminative and reliable information used as the feature descriptor for a video cuboid. The detailed introduction of these transformation techniques is elaborated in this section.

**(1) Discrete Fourier Transform**

In mathematics, Fourier Transform was a revolutionary concept and gave new direction of the development of mathematics. Simply speaking, Fourier Transformation states any function (the domain of this function is called *time domain*) can be expressed as the integral of sines and/or cosines of increasing frequency multiplied by a weighting function (the resulted function is accordingly called *frequency domain*); this function in time domain can be very complicated but it should meet certain mild constraints. The function, expressed in a Fourier transform, can be reconstructed (recovered) completely via an inverse process. These important properties of Fourier transform allow us to work in the 'frequency domain' and then return to the original domain without losing any information. By 'frequency', we mean it is a function of spatial coordinates, rather than time. For example, if an image represented in frequency space has *high* frequencies then it means that the image has sharp edges or details. The continuous Fourier transform is formulated as follows:

$$F(\omega) = \int_{-\infty}^{+\infty} f(t)e^{-i2\pi nt}dt \qquad (2)$$

where *f(t)* is any signal, *F(ω)* is the Fourier transform of *f(t)*, $i = \sqrt{-1}$ and *n* is often called the *frequency variable.*

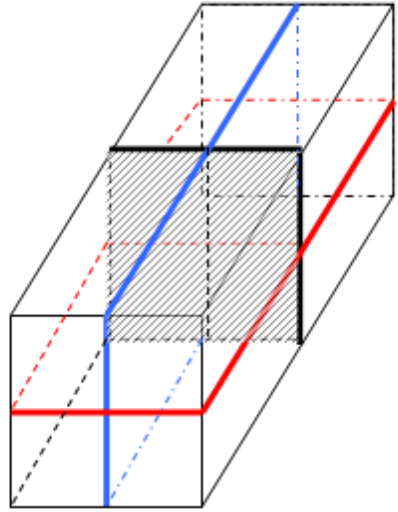

Fig. 10  Illustration of a cuboid and its internal three orthogonal planes.

The Discrete Fourier Transform is a specific Fourier transform which works on the sampled signal function and its non-zero values have a limited (*finite*) duration. Due to this property, DFT is useful in transforming function of discrete values, for example pixel values in images. The definition of DFT is:

$$F_n = \sum_{k=0}^{N-1} f_k e^{-2\pi ink/N} \qquad (3)$$

The complex numbers $f_0$ … $f_N$ are transformed into complex numbers $F_0$ … $F_N$.

In Fig. 11, an example of a discrete function and its Discrete Fourier Transform is given.

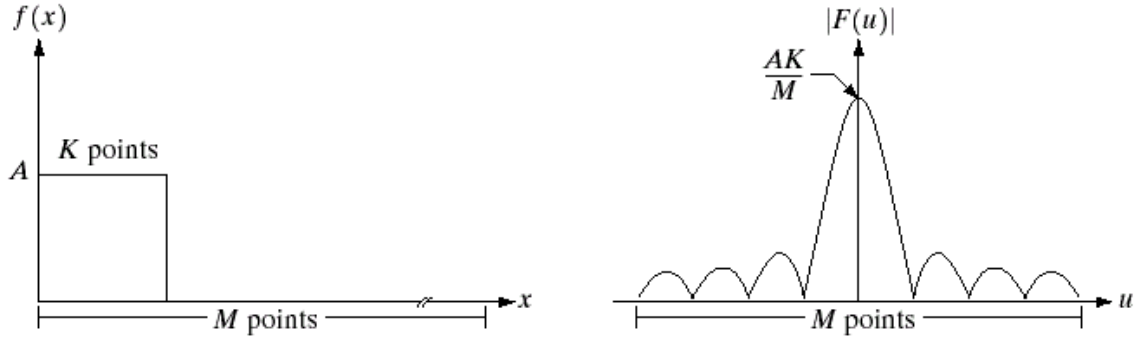

Fig. 11  A discrete function of M points (left) and its Fourier spectrum (right).

In our implementation, the magnitudes of complex coefficients $F_N$ are used as the descriptor of a sequence. Thanks to the periodicity of Discrete Fourier Transform, the descriptor we obtained is insensitive to geometrical changes like rotation, scaling and also the choice of the starting point.

However, since the support of the base function $e^{-i2\pi nt}$ covers the whole temporal domain (i.e. infinite support), $F(\omega)$ depends on the values of $f(t)$ for all times. This makes Fourier transform a global transform that cannot analyze local or transient properties of the original signal $f(t)$. [13]

**(2) Discrete Cosine Transform [16]**

The Discrete Cosine Transform decorrelates the image data, like other transform techniques, to coefficients which are encoded independently for further usage without losing compression efficiency; it is derived from the Discrete Fourier Transform where the sine component of DFT is zero. There are 8 types of DCT of which 4 are common. The most common DCT definition of a 1-D sequence of length $N$ is:

$$C(u) = \alpha(u) \sum_{x=0}^{N-1} f(x) \cos[\frac{\pi(2x+1)u}{2N}] \quad (4)$$

where $u = 0, 1, 2, \ldots, N-1$; and $\alpha(u)$ is defined as:

$$\alpha(u) = \begin{cases} \sqrt{\dfrac{1}{N}} & for \ u = 0 \\ \sqrt{\dfrac{2}{N}} & for \ u \neq 0 \end{cases} \quad (5)$$

Ignoring the $f(x)$ and $\alpha(u)$ of equation (4), we plot the $\sum_{x=0}^{N-1} \cos[\frac{\pi(2x+1)u}{2N}]$ for N=8 in Fig. 12. The first top-left waveform ($u = 0$) renders a constant (DC) value, whereas, all other waveforms ($u = 1, 2, \ldots, 7$) have progressively increasing frequencies. These waveforms are called the *cosine basis functions*. These basis functions are orthogonal which means multiplication of any waveform in Fig. 12 with another waveform followed by a summation over all sample points yields a zero (scalar) value, whereas multiplication of any waveform in Fig. 12 with itself followed by a summation yields a constant (scalar) value. Orthogonal waveforms are independent, that is, none of the basis functions can be represented as a combination of other basis functions.

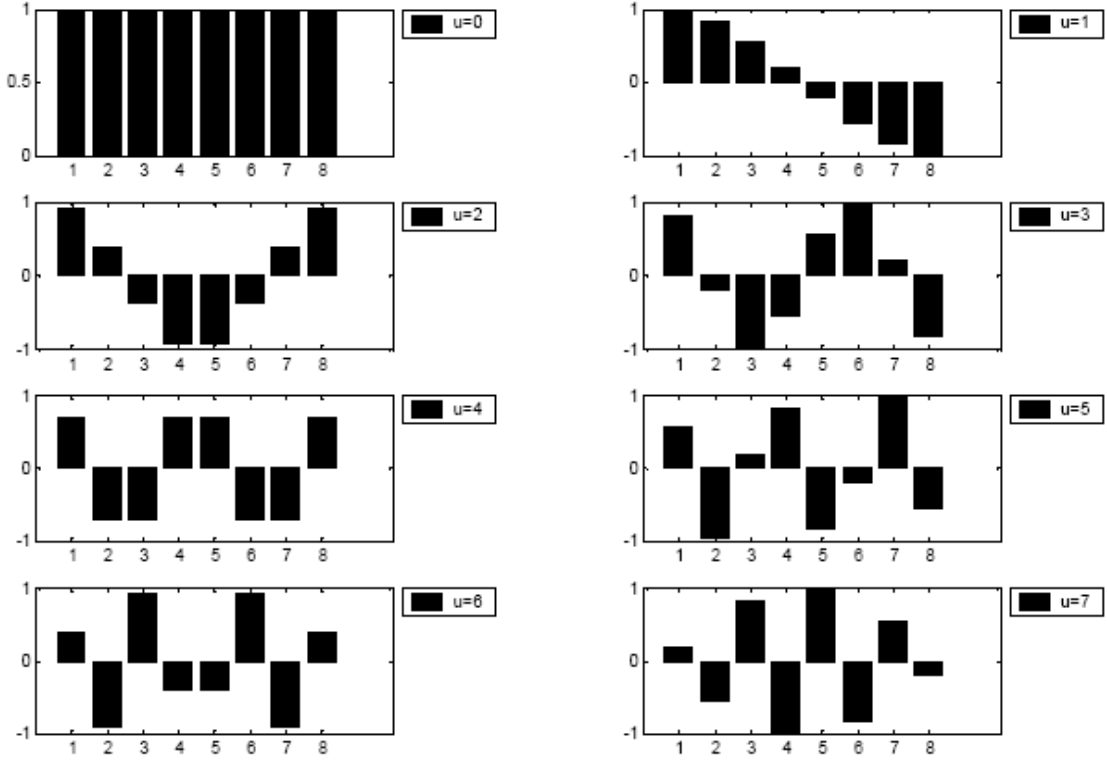

Fig. 12  One dimensional cosine basis function (N=8). [11]

One dimensional discrete cosine transform is useful in processing one-dimensional signals such as speech waveforms. For analysis of two-dimensional (2D) signals such as images, a 2D version of the DCT is required. For a $N \times M$ matrix A, the 2D DCT is computed as an extension of 1D DCT: applying 1D DCT to each row of A and then to each column of the result. Thus, the transform of $A$ is given by:

$$C(u,v) = \alpha(u)\alpha(v)\sum_{x=0}^{N-1}\sum_{y=0}^{N-1} f(x,y)\cos[\frac{\pi(2x+1)u}{2N}]\cos[\frac{\pi(2y+1)v}{2N}] \quad (6)$$

where $u$, $v$ =0, 1, 2,…, $N-1$ and $\alpha(u)$ and $\alpha(v)$ are defined as equation (5).

The 2D basis functions are generated by multiplying the horizontally and vertically oriented 1D basis functions. Fig. 13 gives the illustration of basis functions for N=8 which exhibit progressive increase in frequency in both the vertical and horizontal orientations.

As discussed above, DCT shows many good properties which make it particularly useful in the image processing area. These properties include: a). Decorrelation. It can remove the redundancy between neighboring pixels. b). Energy compaction. DCT packs input data into as few coefficients as possible and discards coefficients with relatively small amplitudes without losing information. c). Orthogonality, it guarantees the independency between two components, that is, the effect of modifying one component will not affect other components.

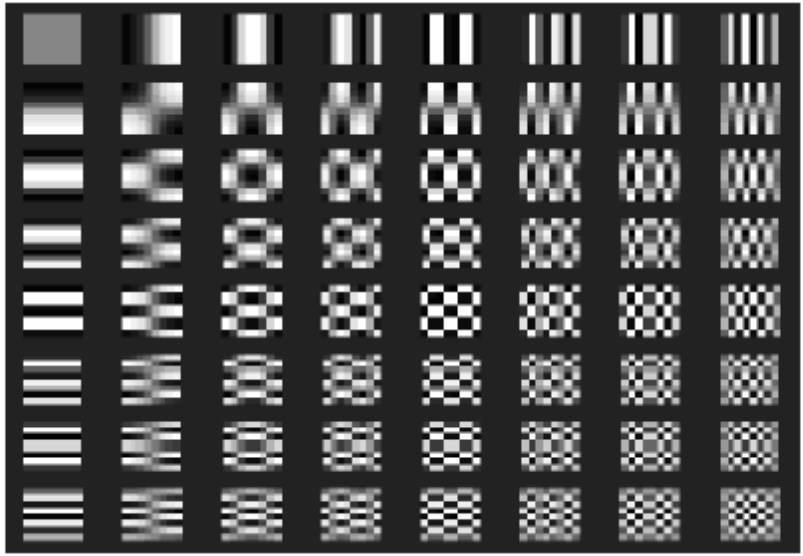

Fig. 13 Two dimensional DCT basis functions (N = 8). Neutral gray represents zero, white squares represent positive amplitudes, and black represents negative amplitude [16].

**(3) Discrete Wavelet Transform [10]**

The development of Wavelet Transform is contributed by a number of researchers. Initially, a French geophysicist, *Jean Morlet*, came up with an ad hoc method to model the process of sound wavelet traveling through the earth's crust. Unlike Fourier analysis, he did not use sine and cosine curves, but simpler ones which he called 'wavelets'. *Yves Meyer*, a French mathematician, recognized this work to be part of the field of harmonic analysis, and came up with a family of wavelets that he proved were most efficient for modeling complex phenomena. This work was improved upon by two researchers in America, *Stephane Mallat* of New York University and *Ingrid Daubechies* of Bell Labs. Since 1988, there has been a small explosion of activity in this area, as engineers and researchers apply the Wavelet Transform to applications ranging from image compression to fingerprint analysis. The Wavelet Transform has even been implemented in silicon, in the form of a chip from Aware Inc.

Fourier Transform can not be used to do the localization analysis, because it only obtains the whole frequency analysis of the signal. However, wavelets are new families of orthonormal basis functions which overcome the problems of the Fourier Transform. Unlike the sine and cosine waves of the Fourier Transform, they need not have infinite duration. They can be non-zero for only a small range of the wavelet function. This manifests its 'compact support' property which allows the Wavelet Transform to translate a time-domain function into a representation that is localized not only in frequency (like the Fourier Transform) but in time as well. This ability has brought forth new developments in the fields of signal analysis, image processing, and data compression.

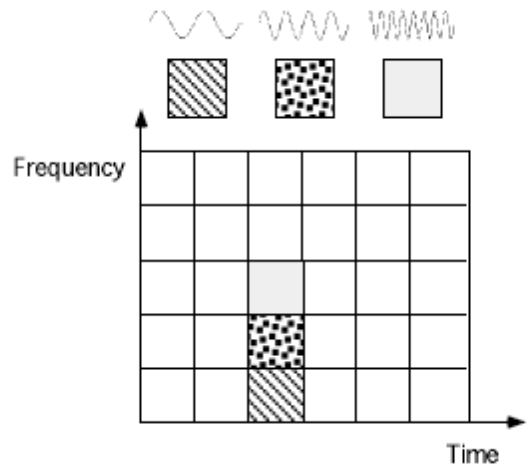

Fig. 14  Fourier basis functions, time-frequency tiles, and coverage of the time-frequency plane. [13]

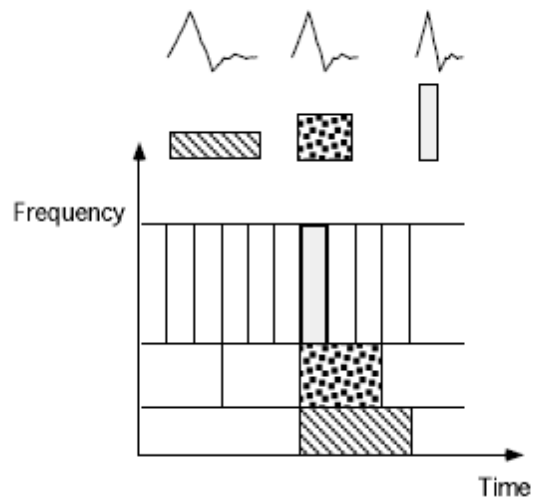

Fig. 15  Illustration of time-frequency tiles, and coverage of the time-frequency plane with Daubechies wavelet basis function. [13]

The time-frequency resolution differences between the Fourier Transform and the Wavelet Transform exists in the basis function coverage of the time-frequency plane [14]. For comparison, we illustrate the Windowed Fourier Transform as well as the Wavelet to explain the difference. Fig. 14 shows a Windowed Fourier Transform, where the window is simply a square wave. The square wave window truncates the sine or cosine function to fit a window of a particular width. Because a single window is used for all frequencies in the WFT, the resolution of the analysis is the same at all locations in the time-frequency plane. [13]

An advantage of Wavelet Transforms is that the windows vary. In order to isolate signal discontinuities, one would like to have some very short basis functions. Meanwhile, in order to obtain detailed frequency analysis, one would like to have some very long basis functions. A way to achieve this is to have short high-frequency basis functions and long low-frequency ones. Wavelet Transforms gives exactly what we expect. Fig. 15 shows the coverage in the time-frequency plane with one Wavelet function: the Daubechies wavelet.

Unlike the Fourier Transform, which utilizes just the sine and cosine functions, Wavelet Transforms have an infinite set of possible basis functions. Thus Wavelet analysis provides immediate access to information that can be obscured by other time-frequency methods such as Fourier analysis.

As explained above, one of the major disadvantages of Fourier Transform is that it decomposes the localized, non-periodic functions into non-localized, periodic functions. This is doing no good when analyzing local information, like discontinuities or silhouette information in certain position. Fortunately, the adoption of multiple basis functions of wavelets gives perfect representation of the local information.

Here the wavelet is explained mathematically. It starts with a mother wavelet, from which all (child) wavelets are created by translation in space and dilation. Let γ(x) be a wavelet (mother wavelet). A set of wavelets can be derived from γ(x): [16]

$$\Upsilon_{a,b}(x) := \frac{1}{\sqrt{a}} \Upsilon(\frac{x-b}{a}), (a,b \in \mathbb{R}, a > 0) \quad (7)$$

Where *a* is the dilation parameter and *b* is the translation parameter.

The Discrete Wavelet Transform is a mapping T: $L^2(\mathbb{R}) \to l^2(\mathbb{Z}^2)$

$$(Tf)_{a,b} = \int_{\mathbb{R}} f(x) \Upsilon_{a,c}(x) dx \quad (8)$$

Where the mother wavelet $\Upsilon_{a,c}(x)$ satisfies:

$$\int_{\mathbb{R}} \Upsilon(x) dx = 0 \quad (9)$$

As all the good properties mentioned above, Wavelet Transform is expected to be quite good in feature extraction. We make use of the resulted frequency coefficients as a compact and discriminative descriptor for a sequence.

### 3.2.4 Correlogram of Oriented Gradient

*Dollár et al.* [6] calculate the 3D gradient for each pixel of a cuboid. Considering the 3 gradients of each pixel, we select a reference gradient ($G_x$) and obtain the ratios of $G_y/G_x$ and $G_t/G_x$. Accordingly the three channels ($G_x, G_y, G_t$) formed by 3D gradient are reduced to be two ($G_y/G_x, G_t/G_x$) which are also the same size as cuboids. This elimination simplifies the formation of correlogram.

We partition $G_y/G_x$ and $G_t/G_x$ equally and obtain certain ratio interval combinations. The number of combination is equal to the square of partition numbers. On the one hand, each histogram bin represents one of this combinations and the value of it is the magnitude summation of each pixel whose gradient ratios satisfy this interval combination. For instance, one histogram bin is to find the pixels whose $G_y/G_x$ belongs to [0.1, 0.2] and $G_t/G_x$ belongs to [0.5, 0.6]. The obtained histograms from cuboids altogether constitute a discriminative representation of one sequence. On the other hand, correlogram is to compute the co-occurrence of pixels for a pair of ratio combinations with regard to distance. In fact, the correlogram of a cuboid is a 3-dimensional table of statistics for ratio combination pairs, of which the *k*-th entry for pair <*i, j*> is the magnitude summation of pixels belonging to ratio combination *i* around a pixel falling into ratio combination *j* at a distance *k*. The formed correlogram is apparently a sparse matrix, and is flattened to be a vector and projected to a lower dimension by PCA.

## 3.3 Classification methods

In the classification phase, we adopt both the widely used SVM classifier and the straight forward KNN in the implementations. As each sequence is represented by a histogram of occurrence of each video word resulted from K-means clustering, thus the length of the histogram is equal to the number of video words, and these histograms built from all the sequences are the input to classifiers. After certain computation, the classifiers will give the predicted class labels that the test sequences belong to.

### 3.3.1 SVM

Support Vector Machine (SVM) is a popular technique for classification. Generally, it is a binary classification algorithm that looks for an optimal hyperplane as a decision function in a high dimensional space [27]. A classification task usually involves with training dataset $\{x_k, y_k\} \in \{-1,1\}$ and testing dataset, where $x_k$ contains the training features and $y_k$ is the class labels. The goal of SVM is to produce a model which predicts the class labels based on the given feature values in the testing set.

In the binary case, the theory behind SVM is to find a linear separating hyperplane with the maximal margin in this higher dimensional space. As shown in Fig. 16, margin is the distance between two hyperplanes defined by support vectors. When the margin reaches its maximal value as hyperplanes adjust, the distance between the middle hyperplane and its nearest point is maximized, and this plane is the hyperplane we expect.

Mathematically, SVM is to solve the following optimization problem:

$$\min_{w,b,\xi} \quad \frac{1}{2} w^T w + C \sum_{i=1}^{l} \xi_i \quad (10)$$

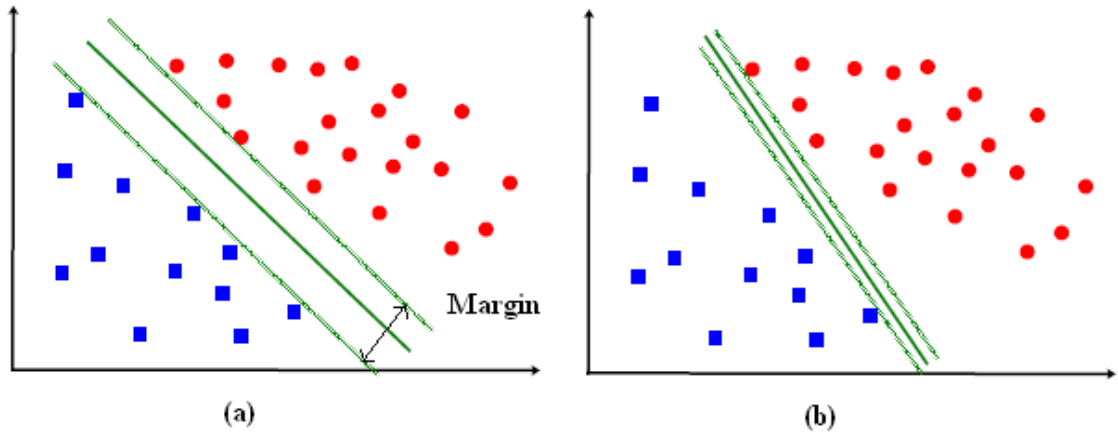

Fig. 16 Two possible hyperplanes for categorizing point set.

$$\text{Subject to: } y_i(w^T\phi(x_i)+b) \geq 1-\xi_i \quad (\xi_i > 0) \quad (11)$$

The solution to this problem is using Lagrangian theory. As the mathematical details are skipped (for more detail refer to [28]), we only give the conclusion of the solution, which is the SVM problem ultimately depends only on the choice of a kernel function. And the four common used basic kernels are: 'Linear', 'Polynomial', 'Radial basis function' and 'Sigmoid' kernel.

Of which Radial Basis Function (RBF) has the following formulation:

$$K(x_i, x_j) = \exp(-\gamma \| x_i - x_j \|^2) \quad (\gamma > 0) \quad (12)$$

RBF kernel is a nonlinear function and can handle the case when the relation between class labels and attributes is nonlinear; and the number of parameter, which affects the complexity of model selection, is less than that of linear kernel functions; moreover, RBF kernel has less numerical difficulties. Due to these good properties, RBF kernel is chosen in our implementation. And the SVM problem squeezes to find the optimal $\gamma$ and $C$ (penalty parameter of error term in equation (10)). The best parameters $C$ and $\gamma$ are not known beforehand, therefore some kind of model selection (optimize parameters for better accuracy in predicting the testing data) must be performed. We select cross validation to achieve this goal. In n-fold cross-validation, the training set is divided into n subsets of equal size. Each subset is tested using the classifier trained on the remaining (n – 1) subsets [26]. Basically, pairs of ($C$, $\gamma$) are tried and the pair with the best cross-validation accuracy is picked. In *Chang et al.*'s program [27], a 'grid search' is implemented in optimizing $C$ and $\gamma$ when running cross validation. They believe that exponentially growing sequences of $C$ and $\gamma$ are practical to identify good parameters (for example, $C = 2^{-5}, 2^{-3}, ..., 2^{-15}, \gamma = 2^{-15}, 2^{-13}, ..., 2^3$).

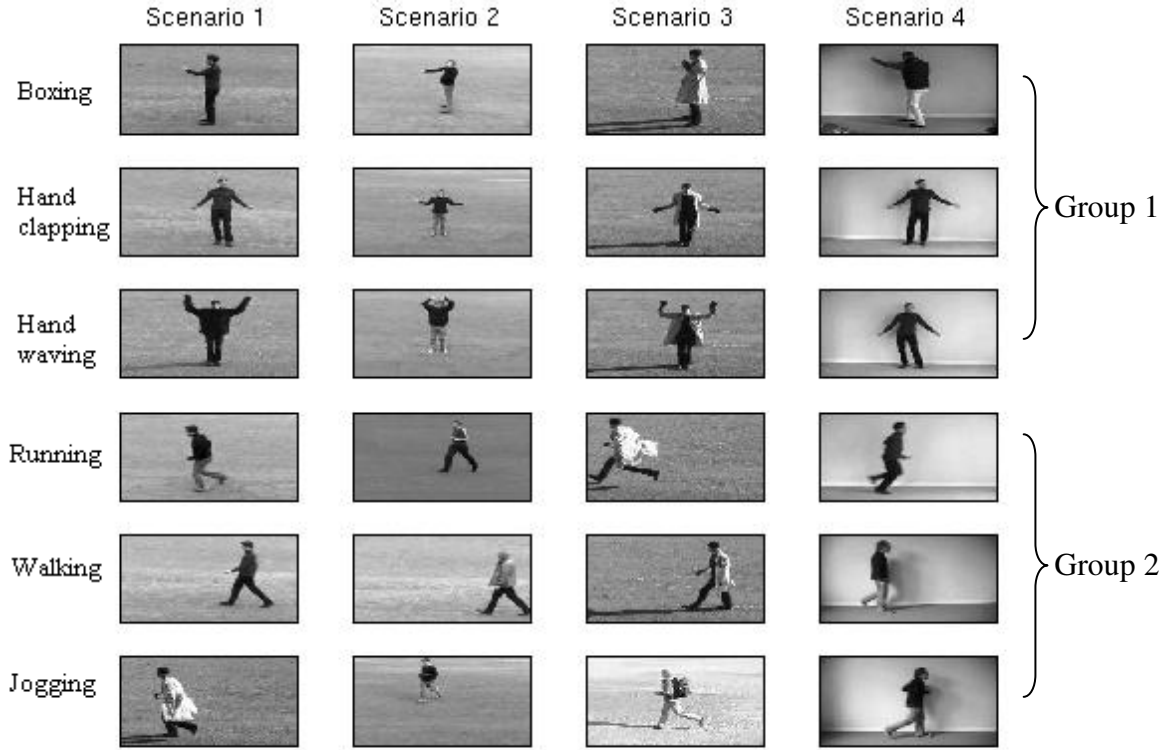

Fig. 17  Example frames of each action of KTH database.

### 3.3.2 KNN

K Nearest Neighbor (KNN) is the simplest and mostly used classifier. It is to assign an object to a class according to the vote of its k nearest neighbors. K is an integer and normally small. If K = 1, the object is directly assigned to the class of its nearest neighbor.

## 3.4 KTH dataset

The KTH human action dataset is a mostly exploited while very challenging dataset. It contains 600 video sequences. Each video has only one action. This dataset consists of six classes of actions (*Boxing, Handclapping, Handwaving, Running, Walking and Jogging*) performed by 25 subjects in four scenarios: outdoors, outdoors with scale variation, outdoors with different clothes and indoors [18]. The environment of the shooting scene is restricted to be as homogeneous (no background noise) as possible. These six classes of actions can be further categorized into two groups: one is actions related to hand motions including the first three actions, while the other group is actions related to leg motions including the last three. See Fig. 17. Actions from the same group are quite similar, sometimes even difficult to discriminate by human eyes, for instance '*Jogging*' and '*Running*'.

For all experiments on this dataset, we follow the same training and testing dataset division: sequences of 6 actions performed by 16 subjects are used for training (totally 384 sequences) and the rest videos (216 sequences) performed by the other 9 subjects for testing. For each sequence, only the first 300 frames are selected because actions are performed periodically.

# 4 Shape context

In this section, we give the implementation details of approaches related to shape context, including the parameters settings and the result comparisons. In order to achieve reliable accuracies, all the presented results are averaged by 20 runs.

## 4.1 Parameter settings

In the detection phase, *Dollár et al.'s* detection method is adopted. The parameters to be set are the $\sigma$ for the 2D Gaussian smoothing kernel and $\tau$ for the quadrature pair of 1D Gabor filters applied temporally. We set $\sigma$= 2.5 and $\tau$ =1.5 for all the experiments using this extraction method. Considering the memory limit and action variations, we select 300 frames from each sequence for interest point detection.

In the description phase, due to the fact that projected 3DSC is making use of the original 2DSC on projection planes, thus the parameters of projected 3DSC are mainly the partitions of radius $r$, and angle $\theta$ and the number of projection plane. For 3DSC, there is one more angular parameter $\varphi$ to be set besides $r$ and $\theta$. For fair comparisons, the number of radial bins $NO\_r_{2DSC}$ is set to be equal to $NO\_r_{3DSC}$, and number of angular bins $NO\_\theta_{2DSC}$ equal to $NO\_\varphi, \theta_{3DSC}$. In our implementation, the maximum distance between any two interest points is selected as the radius of circle or sphere kernel in order to include every point within the kernel. The radius is partitioned to be 10 and 15 segments and the circle or sphere kernel is divided to be 16, 24, 32, and 40 angular bins. Therefore, we obtain 8 kinds of settings by combining different radius and angular partition. For instance, 10 radial bins and 16 angular bins form a projected 3DSC descriptor of length equal to the radial number times angular number times number of projection planes, which is 10*16*3 =480, while a 3DSC descriptor has the length of 10*16*16 =2560, which is mapped to a lower dimension of length 100 using PCA for the reason of computer memory.

Finally, in the classification phase, we adopt non linear Support Vector Machine (SVM) with Radial Basis Function (RBF) kernel and the library libSVM [26] to do the classification. The best parameters $C$ and $log(\gamma)$ are selected by 5-fold cross validation in a grid search on the training data. An example of SVM training process is given in Fig. 18. The other classifier we applied is the KNN using $\chi^2$ distance [6], and K is set to be 5 instead of 1 in all the experiments to eliminate the randomness of classification.

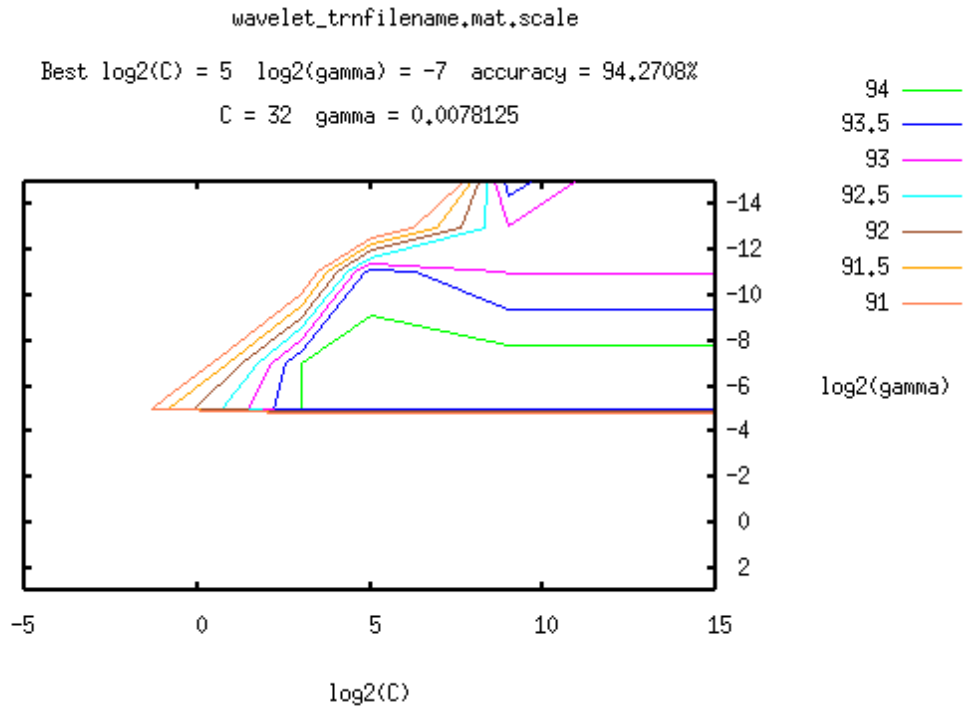

Fig. 18  Example of SVM tuning its parameters.

## 4.2 Experimental results

Fig. 19 displays the results of 3DSC and projected 3DSC under all settings. For 3DSC, the recognition accuracy oscillates between 68% and 71%. The accuracy barely improves as the number of angular or radial bins increase. This is because the distribution of spatial temporal points is already very loose, thus even we refine the partitions in radius or angle, the refinement can not improve the results significantly, and the accuracy is still dominated by the imprecise point distribution. On the other hand, for projected 3DSC shown in Fig. 19 (b), the accuracy increases as the number of angular bins grows, while the number of radial bins provides no acceleration in accuracy. This can be explained as more angular subdivisions in kernel give better discrimination in describing two objects under the condition that these two objects are discriminative enough. Due to the fact that the projected 3DSC manifests more difference between actions than 3DSC, thus the refinement of the number of angular bins in projected 3DSC can improve the recognition accuracy. The obtained results also coincide with the common pattern that coarse spatial sampling and fine orientation sampling turns out to be the best strategy in human detection or representation [38]. In Fig. 20, the comparison of these two approaches has been given under the corresponding settings. Apparently, projected 3DSC outperforms 3DSC by more than 10% in accuracy under all circumstances. And this result testifies the efficiency of the projected 3DSC.

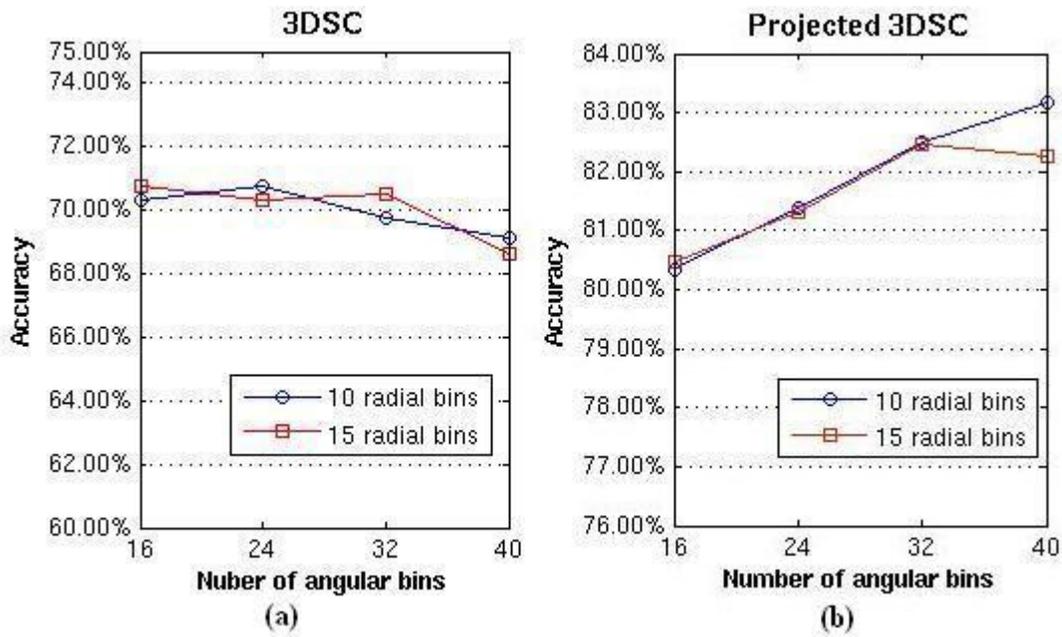

Fig. 19  Result comparisons between 3DSC and projected 3DSC.

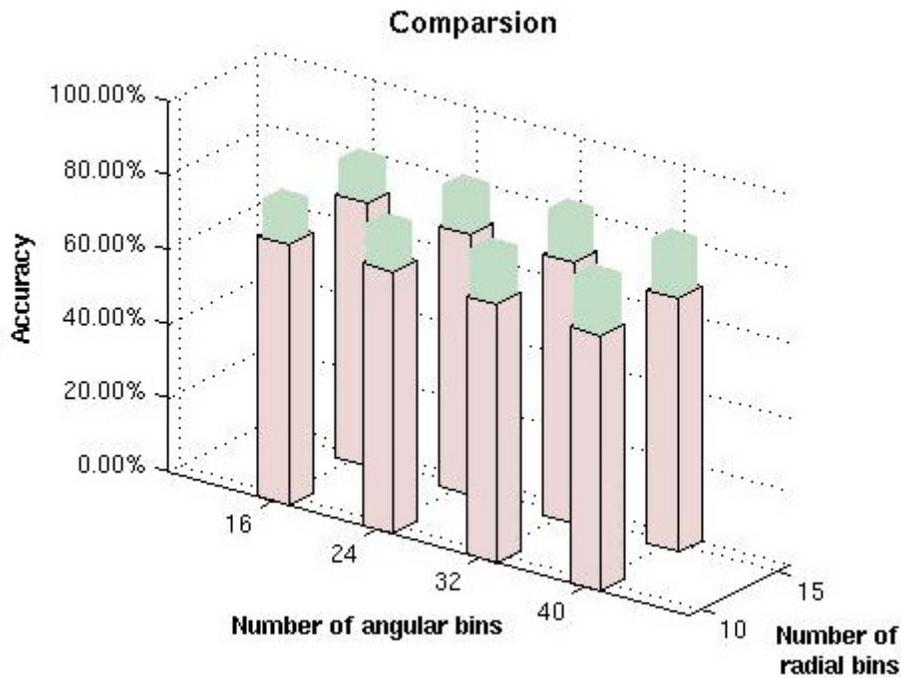

Fig. 20  Accuracy comparisons between 3DSC and projected 3DSC. Pink bars indicate accuracy achieved by 3DSC, and green bars indicate accuracy achieved by projected 3DSC.

The comparison of confusion matrices of these two algorithms is shown Fig. 21. Fig. 21 (a) shows that 3DSC has good performance in *'Handclapping'* and *'Handwaving'* but gives very bad results in the remaining actions. While Fig. 21 (b) shows that projected 3DSC brings great improvements in *'Boxing'* and *'Walking'* by more than 15% compared to 3DSC, and still keeps the recognition rate of the other actions. As we mentioned before, these six actions can be divided into two groups, one is actions related to hands (*'Boxing'*, *'Handclapping'*, *and 'Handwaving'*) and the other is actions related to legs (*'Running'*, *'Walking'*, *and 'Jogging'*). As illustrated in Fig. 21 (a), some sequences belonging to group 1 are misclassified to group 2 (For example, 3% of *'Boxing'* are classified as *'Jogging'*); similarly, some of the group 2 are misclassified to group 1(For instance, 4% of *'Running'* are categorized to be *'Handwaving'*). However, projected 3DSC greatly eliminates this inter-group ambiguity, as shown in Fig. 21 (b). This conclusion proves projected 3DSC's efficiency in discriminating actions within each group, as well as eliminating misclassification between groups.

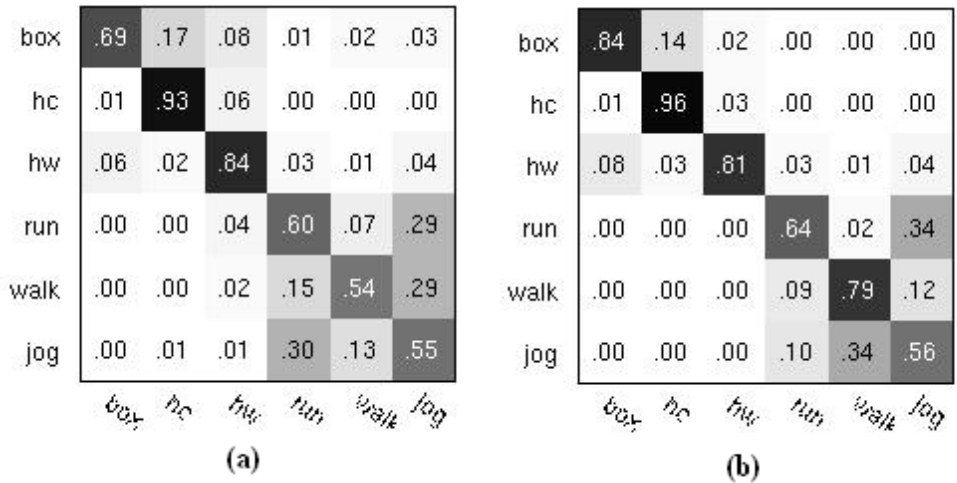

Fig. 21  Confusion matrices of 3DSC (a) and projected 3DSC (b).  The settings of these results are: radius is divided by 10, and number of angular bins is 20.

# 5 Transform based descriptors

In this section, the experimental results of the transform based descriptors as well as the comparisons between them are illustrated; meanwhile, possible explanations are also given to the comparison results.  In order to achieve reliable accuracies, all the results presented are averaged by 20 runs.

## 5.1 Parameter settings

In feature detection and classification steps, we follow the same parameter settings as described in Section 4. In the feature detection step, the cuboids are extracted around interest points, which are the carriers for all the transformation operations as well as gradient calculation.

*3D Gradient:* the spatial temporal gradients are generated within each cuboid and concatenated along three dimensions. The size of the concatenated vector is equal to the number of pixels in the cuboid times the number of smoothing times the dimensions of gradient. Each cuboid, which is approximately six times the scales $\sigma$ and $\tau$ along each dimension, has the length of 17*17*11 = 3179. Therefore in our implementation, the length of descriptor vector is 3179 times 1 times 3 as we only use one smoothing. Then the descriptor vectors of length 9537 are projected to a lower dimensional space of 100 dimensions.

*FFT:* within each cuboid, we extract three orthogonal planes all intersected in the very middle point of the cuboid. Refer to Fig. 10. For each plane, the FFT is applied on it and generates certain number of complex coefficients. The magnitudes of the complex coefficients of the three planes are concatenated together as a descriptor with length of 576.

*DCT:* the same process as Fourier Transform is adopted for DCT, and the resulted descriptors are also of length 576.

*DWT:* we apply Daubechies wavelet transform on these three orthogonal planes. Thanks to its low redundancy, the number of coefficients is the smallest, which is only 288 and less than one thirtieth of the original 3D gradient descriptor.

## 5.2 Experimental results

*Dollár et al.*'s cuboid extraction method is widely used due to its high accuracy, simplicity and fastness. All the experiments in this section are implemented with their extraction method. Fig. 22 shows the accuracy curves regarding to different given conditions. In Fig. 22 (a) and (b), dependencies between accuracy and the number of cuboids detected are illustrated, while relations between accuracy and codebook's size are shown in Fig. 22 (c) and (d). Balancing the accuracy and computational efficiency, we fix the codebook's size to be 750 in Fig. 22 (a) and (b); and set the number of cuboids to be 100 in Fig. 22 (c) and (d). Due to the fact that different description methods generate description vectors of varying length, descriptors of original dimension and reduced dimension are both evaluated in order to achieve fair comparison. The reduced dimension is 100 achieved by PCA.

In Fig. 22, graphs in the first row illustrate the performance comparisons among transform based description methods, of which Fig. 22 (a) displays the results classified by KNN and Fig. 22 (b) using SVM. Apparently, the wavelet based description method outperforms both the FFT and DCT based methods and the best result achieved by Wavelet based method is 93.89% with 150 cuboids and 750 video words. This is because wavelet transform makes use of multiple time-frequency basis functions to capture both the signal discontinuities (short high frequency basis function) and the detailed information (long low frequency basis function). This property guarantees the resulted coefficients bearing intact and precise information extracted from the input imagery. Benefiting from the non-redundancy of Wavelet coefficients, the descriptor based on Wavelet is half the size of those based on Fourier transform and DCT. From Fig. 22 (a) and (b), we can further conclude that the recognition accuracy increases as the size of cuboids grows. This tendency complies with our intuition that more cuboids bring more information from each action, and accordingly enrich the discrimination between two sequences. The best accuracy is achieved with the most cuboids which is 150 for all three transform based description approaches, except for DCT classified by SVM. SVM classifier gives better recognition accuracy than KNN, and for wavelet based description method, the results classified by SVM outperform the results of KNN around 4% regarding to all cuboids numbers.

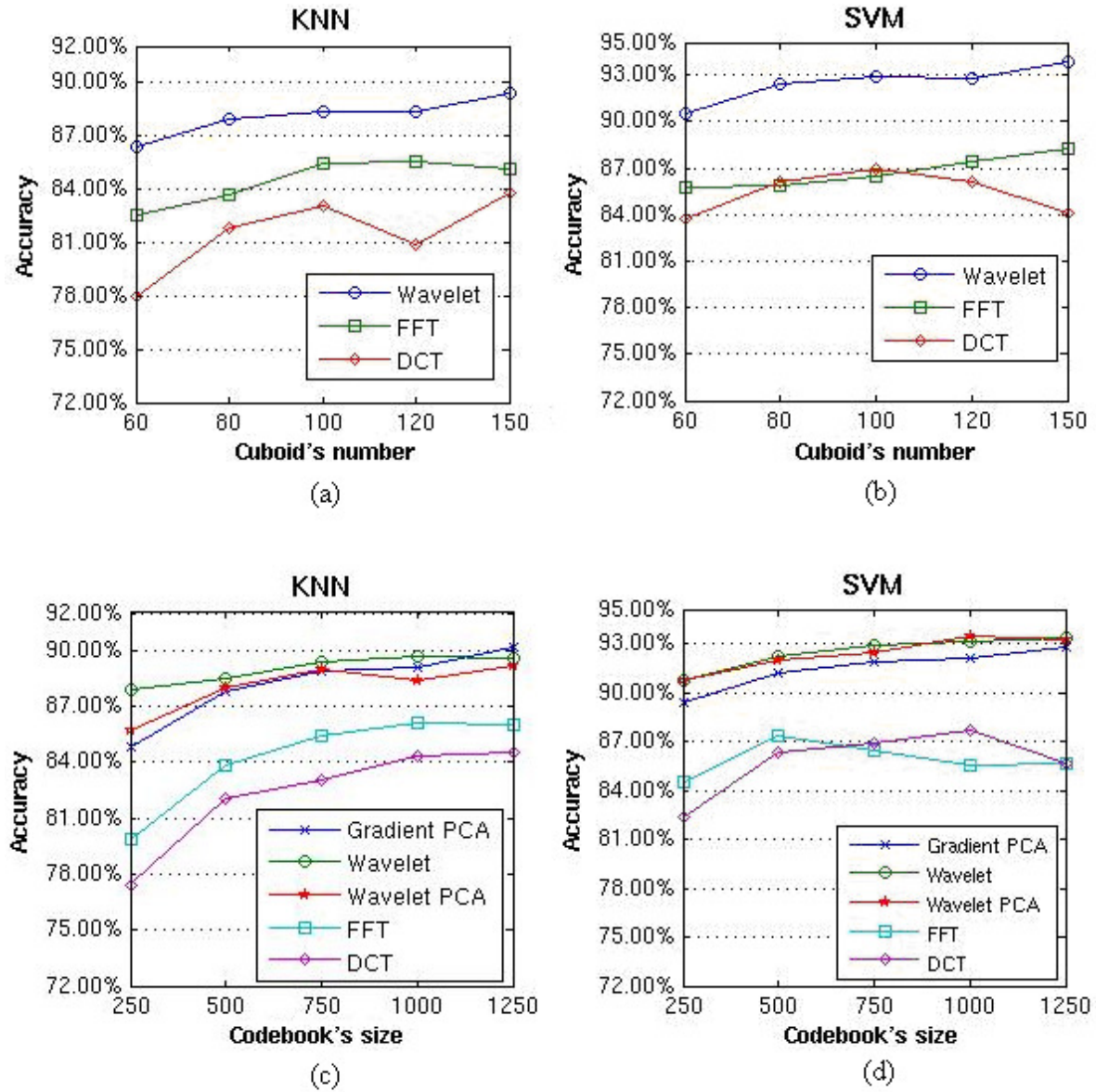

Fig. 22  Comparison of accuracies under different conditions. (a) and (b) show accuracy against cuboids' number given codebook's size equals to 750. (c) and (d) give accuracy against codebook's size given cuboids' number equals to 100.

Fig. 22 (c) and (d) display the comparison of *Dollár et al.*'s 3D gradient description method and the transform based approaches given different codebook' sizes. According to our experimental settings, the 3D gradient descriptor has length of 9537 which encodes extremely rich information. However, Wavelet based descriptor is only with a length of 288, less than one thirtieth of the 3D gradient descriptor. With such a short descriptor length, Wavelet based description still achieves better performance than the 3D gradient. For instance, Wavelet achieves 93.40% in recognition accuracy while 3D gradient obtains 92.75% under the same settings of 1250 video words and 100 cuboids. These results are better than the state-of-the-art solutions, and prove the efficiency and reliability of the coefficients generated by the Wavelet Transform. The dimension of Wavelet based descriptors is reduced to be 100, and it performs nearly the same with that of the original length. This is because the wavelet based descriptor is already very compact and informative, and produces no big difference in performance after PCA, and still outperforms 3D gradient with PCA.

Moreover, Fig. 22 (c) and (d) also give the comparisons among transform based description methods with regard to different codebook's sizes. The size of codebook is an input parameter for K-means clustering. If the size is small, the clustering of these feature descriptors is quite coarse; as this number increases, K-means gives more accurate clustering and manifests more subtle difference between clusters. That is to say, more video words can refine the representation of a sequence better. This conclusion has been verified by the results in Fig. 22 (c) and (d), which show the accuracies of different description methods increase gradually as the size of codebook grows. In addition, the results classified by SVM display overall better performance than KNN.

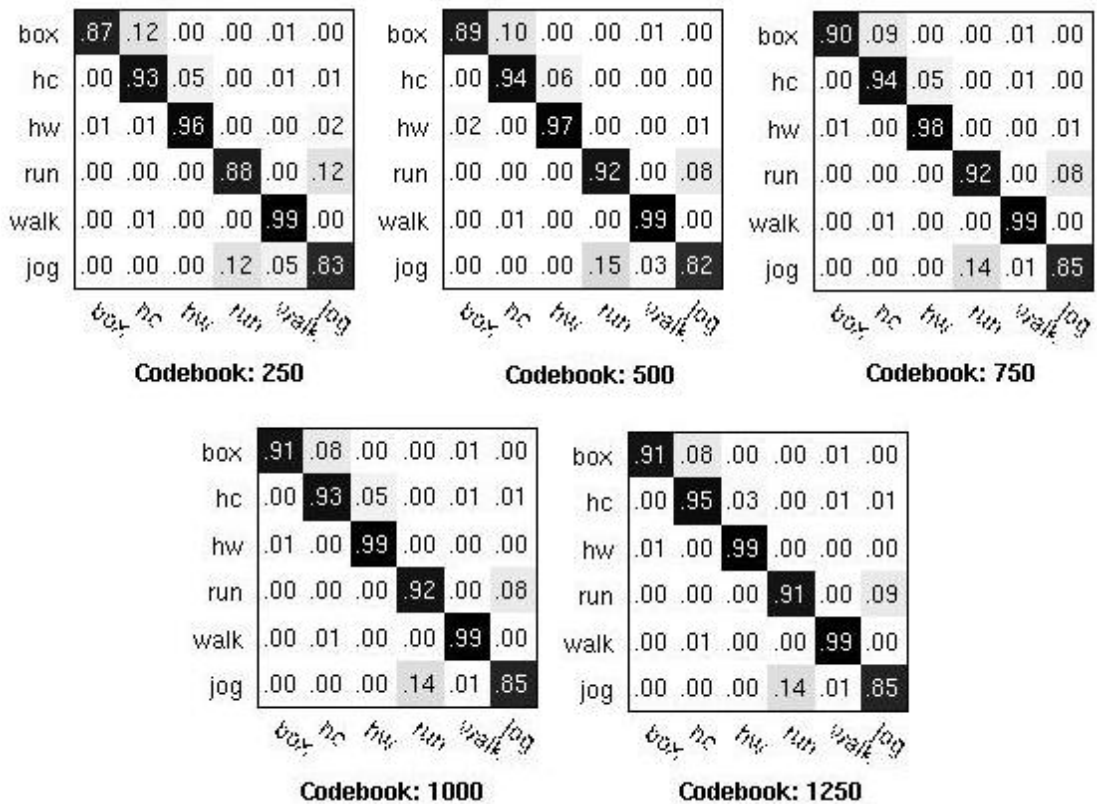

Fig. 23  Confusion matrices of wavelet description using different codebook size.

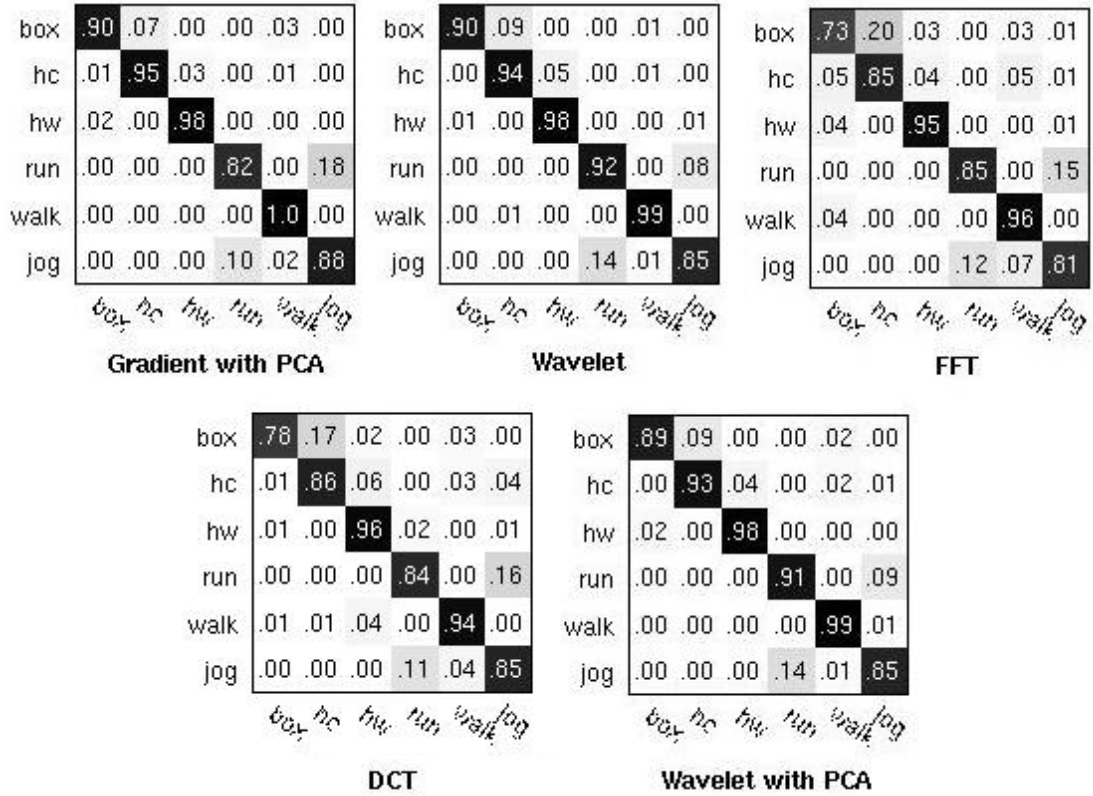

Fig. 24  Confusion matrices of different methods. The settings of all the methods are the same (Sigma and Tau are 2.5 and 1.5 respectively, codebook's size is 750 and number of cuboids is 100).

Different description methods give different recognition accuracies for each of these six action types. Take the Wavelet based description method for example; its recognition accuracies are illustrated in Fig. 23 under different codebook's sizes. The accuracy for each action grows gradually and finally reaches its highest value when the codebook's size is 1250. Meanwhile we demonstrate the recognition capability of different description methods by confusion matrices in Fig. 24. It is clearly indicated that 3D gradient has very good recognition capability in '*Boxing*', '*Handclapping*', '*Handwaving*', and '*Walking*'. But for '*Running*' and '*Jogging*', the results are slightly worse. That is because in the KTH database, some '*Running*' and '*Jogging*' sequences are very similar, and even difficult to distinguish by human eyes. In this case, the 3D gradient is not strong enough to discriminate these two actions. However, the Wavelet based descriptor supplements the accuracy for these two actions. It raises the accuracy of '*Running*' by 10% comparing to 3D gradient while still maintains the recognition accuracy for other actions. In conclusion, the Wavelet based description method is more discriminative to small variations and more capable of stressing the discrimination between actions. In addition, the FFT and DCT based descriptors have similar accuracy for each action and both give bad accuracy rate in '*Boxing*'.

From the comparison results in Fig. 24, we learn that even if Wavelet, FFT and DCT are all widely used in the image processing area, the performance of Wavelet is much better than the other two, meanwhile the descriptor length of Wavelet is just half of that of FFT and DCT. This can be explained by the following outstanding properties of Wavelet: 1). perfect reconstruction ability. This property guarantees no information lost and no redundant information included. That is because the set of basis functions for signal decomposition are varied to accommodate the information extraction. 2). Orthogonality. This property means the independency between two vectors, and when one vector is modified, the other one receives no side effect from the modification. Orthogonality provides orientation selectivity for image processing, and this selectivity is perfectly suitable for human visual system. Detailed explanation is referred to *F. Campbell* and *J. Kulikowski*'s visual system experiment [17].

# 6 Correlogram of Oriented Gradient

In this section, the implementation details as well as the result comparisons of histogram and correlogram are discussed. The results are averaged by 20 runs to get reliable accuracies.

## 6.1 Parameter settings

The detection and classification phases follow the same parameter settings as illustrated in Section 4. And the parameters of the description phase are mainly related to the construction of histogram and correlogram from each cuboid.

*HOG:* the two ratio channels ($G_y/G_x, G_t/G_x$) are equally partitioned into 10 and 12 intervals, thus the lengths of formed histogram are 10*10 = 100 or 12*12 = 144 accordingly.

*COG:* according to the definition in Section 3, $k$ is the distance options and is set to be 3 and 4, and $i$ and $j$ of pair <$i, j$> indicate the ratio combinations. In correspondence with the histogram settings, we adopt the same ratio partition numbers and get two ratio combinations with lengths of 100 and 144. Combining the distance parameter, the length of the descriptor is 100*100*3 = 30000, 100*100*4 = 40000 or 144*144*3 = 62208. To accommodate the limit of computer memory, we reduce the dimension of these vectors to be 100 using PCA.

## 6.2 Experimental results

Fig. 25 displays the recognition results of HOG and COG with regard to the codebook size. The 'blue' and 'green' curves illustrate the performance of HOG with partitions of 10 and 12. Comparing the two curves, we can conclude that 10 or 12 partitions makes no big difference in accuracy. This is due to the fact that histogram is capturing the general distribution, and small refinement of the histogram bins will not improve the final result significantly.

The remaining three curves ('red', 'light blue' and 'purple') show the performance of correlogram against codebook's size. They represent results generated by 3 parameter settings combined by partitions and distances. Compared to the 'red' curve (par: 10, dis: 3), the 'light blue' curve (par: 10, dis: 4) shows almost the same accuracies; while the 'purple' curve (par: 12, dis: 3) manifests big improvement, especially when the codebook size is small. These results comply with our mentioned rule that fine orientation sampling and coarse spatial sampling are the best strategy.

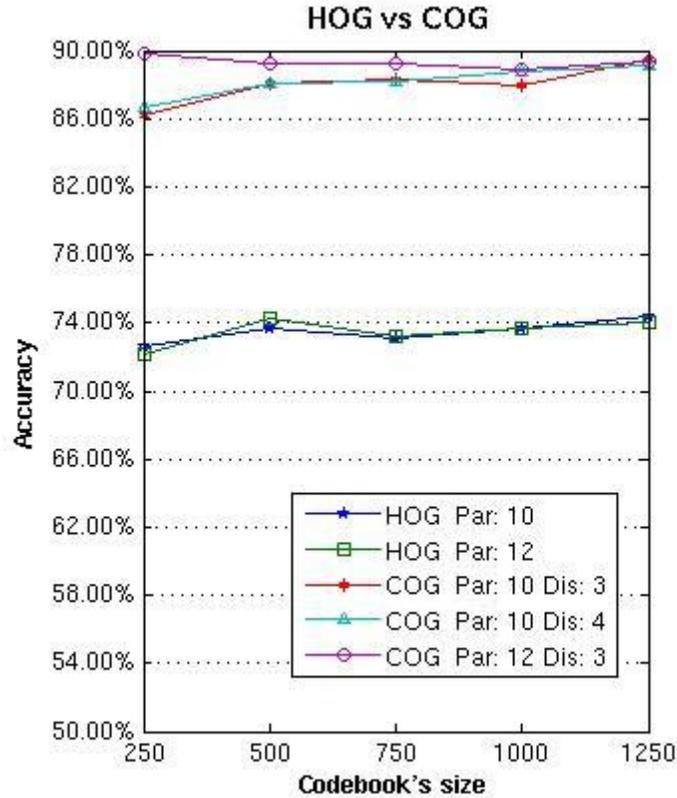

Fig. 25  Comparisons of HOG and COG.

Apparently, COG achieves about 15% improvement over HOG in accuracy under all parameter settings. This significant improvement strongly testifies the discrimination power of the COG descriptor benefiting from its usage of both structural and appearance information.

As shown in Fig. 26, COG and HOG produce better accuracies for '*Handwaving*' and '*Walking*' than other actions, and this outcome reflects the commonness of HOG and COG, while COG displays much better accuracies for each action.

# 7 Further applications in realistic movies

## 7.1 Problem extension

To verify the efficiency of the Wavelet based description method in representing human actions, we apply it to the more challenging Hollywood dataset [25] which has realistic and complex scenarios. Fig. 27 shows some example frames of this dataset. The Hollywood database is composed of action sequences selected from several real movies, such as *'Forest Gump'*, *'Casablanca'* etc. These sequences vary in shooting environment, shooting viewpoint, as well as film editing techniques etc., and moreover, intra-class variations, partial occlusions, photometric variations, camera motion and zooming also create obstacles in the recognition process and hinder the final recognition accuracy. Although some researchers have exploited this challenging dataset, and presented fundamental progress on this issue [25], the existing approaches still give very low accuracy due to these inevitable difficulties of realistic human action datasets, even if they have achieved very good performance in controlled and simplified databases, such as the KTH dataset.

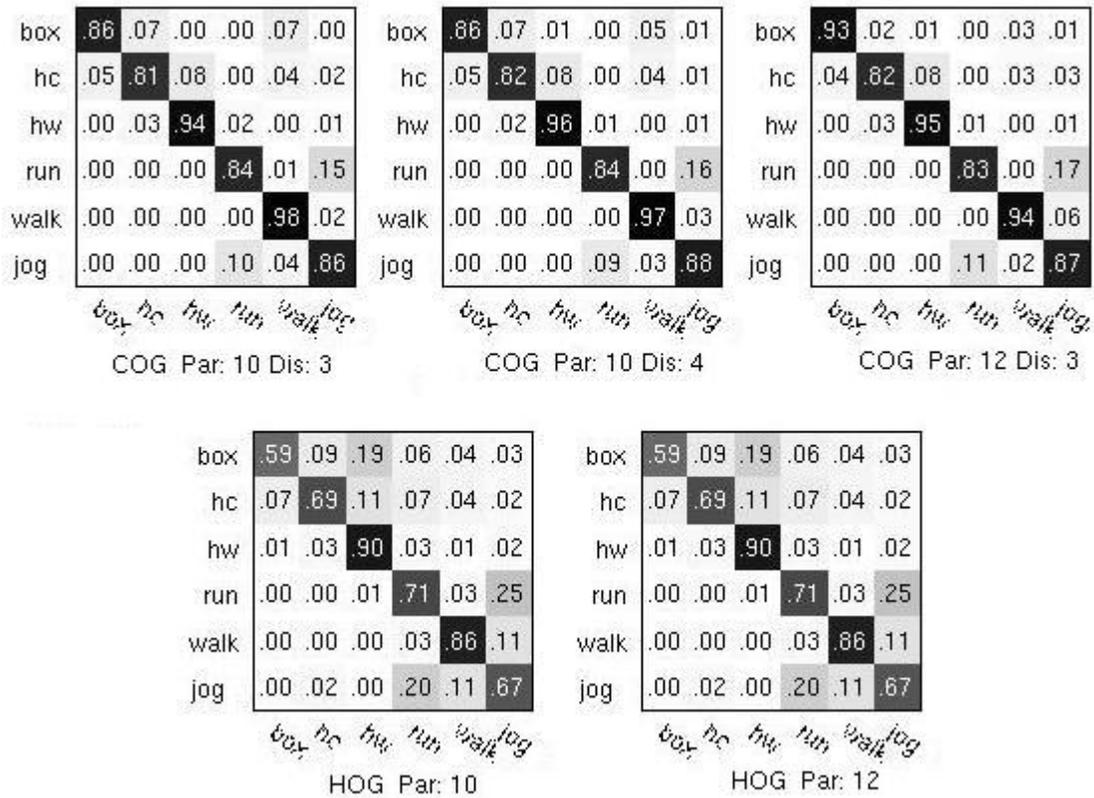

Fig. 26  Confusion matrices of COG and HOG under different parameter settings. The codebook's size is set to be 1000.

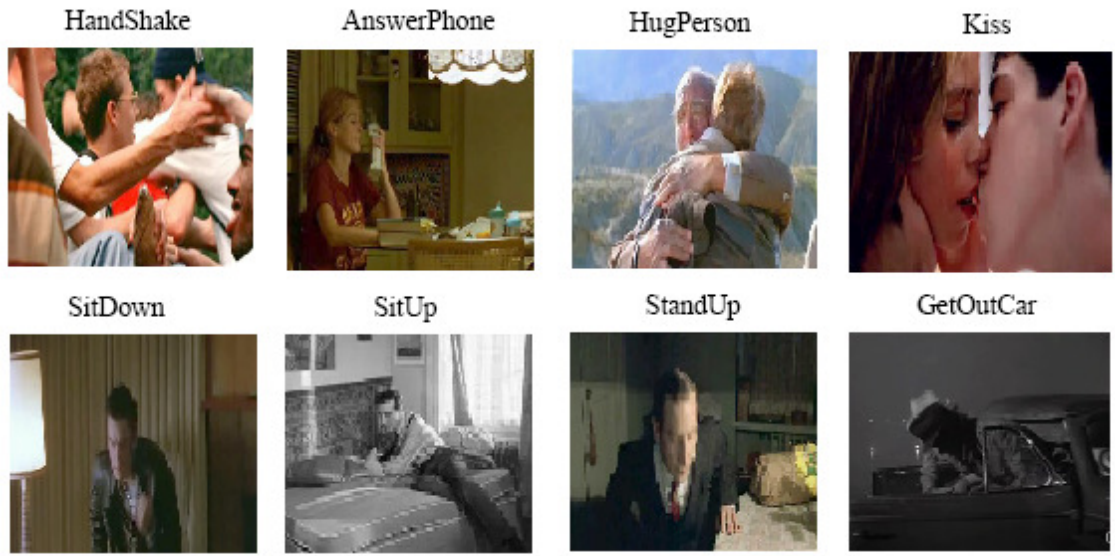

Fig. 27  Example frames of different actions of Hollywood database.

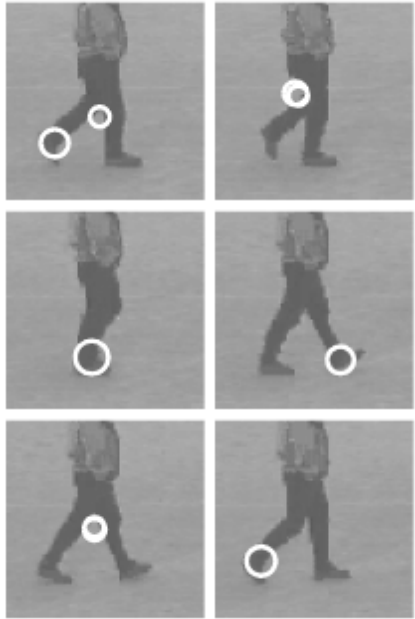

Fig. 28  Results of detecting spatial-temporal interest points from a walking person. [5]

## 7.2 Approaches

We attempt to bring the Wavelet based description method as well as *Dollár et al.*'s 3D gradient to the realistic human action recognition problem and evaluate their performances.

All the realistic movies are processed following the same procedure indicated in Fig. 3. Furthermore, *Laptev et al.*'s extraction method [5] is also implemented in the detection phase as a benchmark approach besides *Dollár et al.*' method [6]. *Dollár et al.*'s method relies on the usage of separable linear filters to obtain the response function of each sequence. Local maxima of the response function are believed to be body parts with strong motions. However *Laptev et al.*'s method is to detect local structures in space-time where the image values have significant local variations in both space and time based on the idea of the *Harris* and *Förstner* interest point operator. In Fig. 28, the detected interest points are shown in a walking person using *Laptev et al.*'s method. In addition, *Laptev et al.* develop a mechanism for selecting spatial temporal scales automatically and the selected scales are roughly corresponding to the size of the detected events in space and to their durations in time [5]. This mechanism facilitates the process of dealing with sequences shot in multiple zooms. As shown in Fig. 29, by using multiple scales, the two '*Kiss*' actions can be detected from these two sequences and can be further matched. Otherwise, we may lose the correspondence between these two '*Kiss*' sequences if only one single scale is adopted.

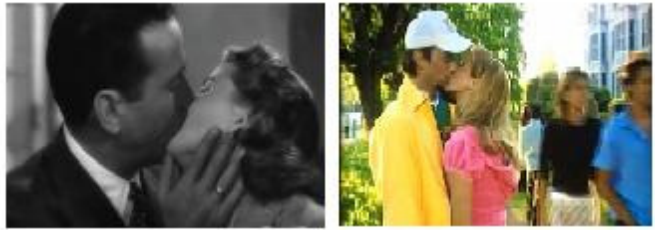

Fig. 29  Action 'Kiss' happens in two scenes with different zooming.

## 7.3 Experimental results

Since all the sequences are clipped from realistic movies, they have varied volumes, some of which are as large as 100 MB, while some are just 130 KB. That is why we select videos with small volumes manually to accommodate the computer memory. Meanwhile, our study on the Hollywood dataset is just a preliminary work, because these is no enough time to preprocess all the sequences, for instance, clipping the actions of interest from large volume sequences. Based on the above difficulties, we select 3 action classes ('*Kiss*', '*HandShake*', and '*StandUp*') and 14 video sequences for each action from the Hollywood dataset. Each sequence is limited to no more than 400 frames presenting activities. In total, the subset consists of 42 video sequences, of which 8 sequences of each action class (24 sequences in total) are used to build the model in training phase, and the remaining 18 sequences for testing.

The parameters of our implementations are tuned to be relatively the same. For *Dollár et al.*'s extraction method, the number of cuboids is fixed to be 100, Sigma is 2.5, and Tau is 1.5. Under this parameter setting, the Wavelet based descriptor has a length of 288, while 3D gradient is 9537 in length and is further projected to 100 dimensions. Fig. 30 (a) gives the comparison results of these two description methods, and apparently, the Wavelet based descriptor outperforms 3D gradient by almost 10%. This validates the efficiency of the Wavelet transform in representing features as well as its superior feasibility in complex movies.

For *Laptev et al.*'s extraction approach, we release the threshold to be e-8 to make sure at least 100 interest points are extracted from each sequence. For some sequences with small number of frames, *Laptev et al.*'s method extracts barely enough points, while for other sequences with more frames, it can detect more than 4000 points using the same threshold, which explains why this extraction method is very time consuming.

All the points are sorted by the response value and the first 100 points are selected as interest points. Accordingly, cuboids are also extracted around each interest point with the same size of 17*17*11 as that of *Dollár*'s method, and the remaining steps are the same as mentioned above. Fig. 30 (b) shows the comparison results using *Laptev et al.*'s extraction method.

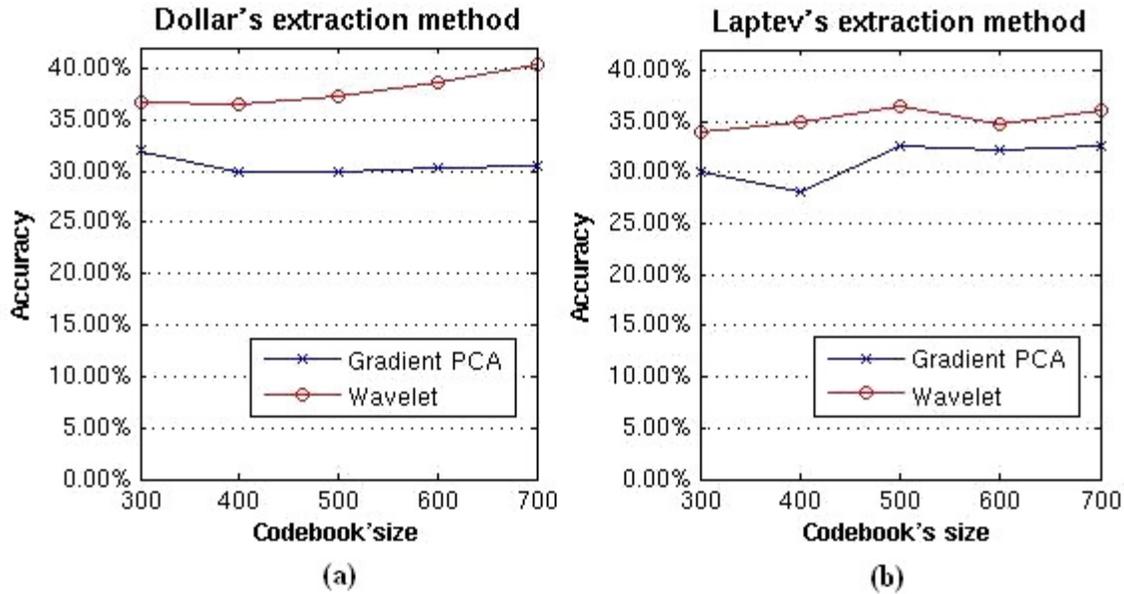

Fig. 30  Comparisons of description methods using different extraction approaches.

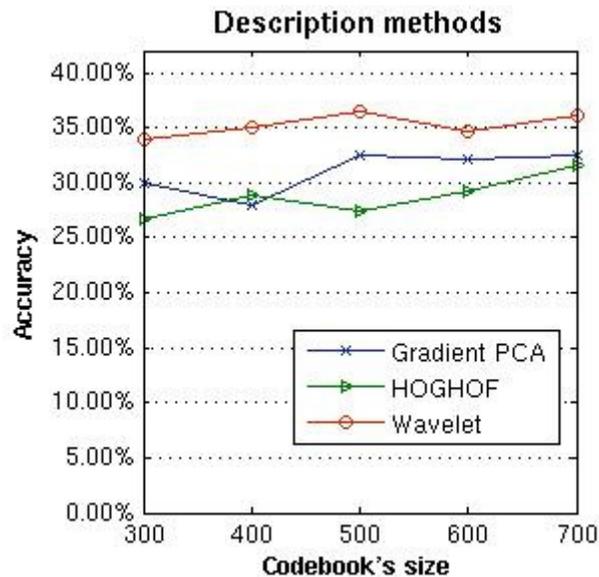

Fig. 31 Accuracies of different description methods we applied on Hollywood database.

Compared to *Dollár et al.*'s method, *Laptev et al.*'s method performs slightly worse. This is because *Laptev et al.*'s method is more focused on the capture of the spatial temporal corners, and often a lot of irrelevant points are detected from the background objects with small motions. This characteristic affects the accuracy of interest point detection and decreases the final recognition accuracy accordingly.

In this thesis, three description methods are applied on the subset of the Hollywood dataset. They are 3D gradient, Wavelet based technique and HOG-HOF (Histogram of Oriented Gradient and Histogram of Optical Flow) [25], respectively. For 3D gradient and the Wavelet based technique, they follow the same setting as mentioned above. While for *Laptev et al.*'s HOG-HOF, we use the executable files downloaded from *Laptev*'s website [25]. HOG is to divide each cuboid into several sub-regions and then build the histogram of gradient directions for pixels of each sub-region, while HOF is similar to the idea of HOG and the only difference is optical flow instead of gradient is histogrammed. And the combination of HOG and HOF generates the HOG-HOF descriptor with a length 162, 72 for HOG and 90 for HOF.

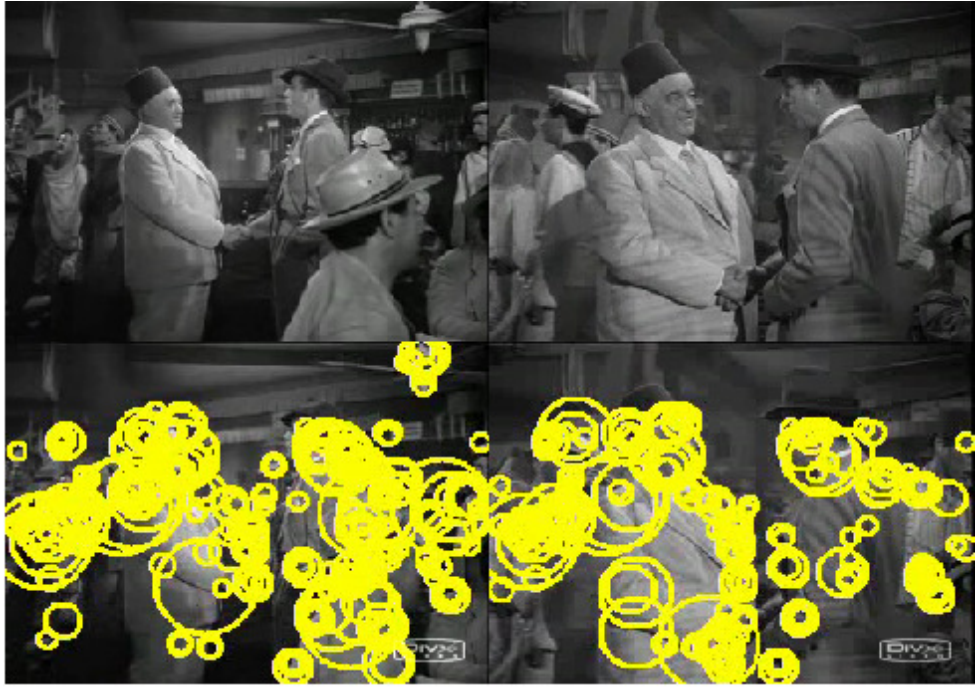

Fig. 32  Two frames transited by a hard cut.

The accuracies achieved by the three description methods are illustrated in Fig. 31. Wavelet based method still gives the best results, while HOG-HOF performs the worst, which is even worse than the 3D gradient. This can be explained by how much information contained in each descriptor. The 3D gradient uses all the original gradient values of each pixel in the cuboid, thus it contains the richest information, while the HOG-HOF distributes all the gradients or optical flow information into each histogram bin, and is inevitable to lose the knowledge about the original distribution. The elimination of structural information in HOG-HOF makes it less discriminative than the other two descriptors.

Conclusively, the accuracy for the Hollywood dataset is significantly lower than that of the KTH dataset. There are 2 major reasons for this bad performance. The first one is because the Hollywood dataset is collected from realistic movies which have more complex scenarios (for instance camera motions, viewpoint change, and zooming etc.) and variations in subjects' actual performing situations (such as partial occlusions, irrelevant extra motion in the background, etc.). The other important obstacle is the scene translation from one frame to another in real films. When a scene is over, the next scene is followed by some translation techniques, such as 'fade in', 'fade out' (often seen in old movies) and 'hard cut' (common in modern movies). This scene translation causes abnormally huge number of interest points detected on these two frames, as shown in Fig. 32. The points indicated by yellow circles are inaccurate and redundant, therefore interfere with the whole interest point detection process, and accordingly affect the recognition results.

# 8 Conclusion and future work

Three novel description methods, which take advantage of the structural distribution and appearance information, are proposed in this thesis, and the implementation details are elaborated as well. The performances of each method along with the mutual comparisons exhibit their reliability and efficiency in human action recognition, and also indicate promising potential to study further on these ideas. Meanwhile, they also imply certain deficiencies and inherent limitations which need special attention in future work.

## 8.1 Shape context

The novel algorithm, projected 3DSC, is proposed to conquer the deficiencies of 3DSC. This approach exploits the structural distribution of interest points projected on the three orthogonal planes, and displays its advantage in human action recognition over 3DSC by specifying the commonness and difference between actions.

However, the projected 3DSC is still incapable of dealing with viewpoint variations. The solution to this issue could be initiating planes intersected by multiple angles (smaller than 90 degree) instead of 90 degree, and projecting point clusters to these planes. In this way, the projection on these chosen planes can be considered as viewing the interest point cluster from different viewpoints, and obtains even more information for further classification.

On the other hand, shape context approaches only encode the structural information of actions, thus it is very difficult to improve the accuracy greatly. One possible way is to refine the partitions of angles, but it can weaken the computational efficiency as the angular bins increase. The other way is to apply different kernels to obtain correlogram description of interest points instead of histogram for more discriminative information.

## 8.2 Transform based descriptors

In this part, transform domain techniques are brought to the unsupervised human action recognition field as the description approaches. These description methods generate promising performance on the KTH human action dataset, and the Wavelet Transform based method outperforms the state-of-the-art recognition algorithms. Moreover, the results indicate that the Wavelet Transform is quite promising and convincing for future usage in realistic human action recognition as it extracts compact and reliable feature information.

These transformation techniques are making use of appearance information extracted from cuboids, hence, one supplementary way in raising accuracy is to combine some structural information of each action by means of descriptor fusion, for instance combining the projected 3DSC. Meanwhile, improvements in the performance of clustering methods and classification algorithms are also desirable. Some upgraded K-means algorithms, for instance approximate K-means [29] which is faster and more accurate than the original one, already exist. And more complex kernel functions can be adopted to improve the SVM classifier.

## 8.3 Correlogram of Oriented Gradient

The histogram and correlogram of spatial temporal gradients are demonstrated, and the results strongly prove that correlogram is much more effective in feature extraction than histogram, due to the fact that correlogram incorporates both the appearance and structural information. However, our implementation of HOG is the most straightforward way, and a lot of work has been done on enhancing its performance by applying histogram on continuous overlapped sub-regions, as well as using detection window and sub-region normalization etc. [25, 38]. These supplements make histogram much more reliable in feature extraction with the expense of algorithm complexity. Apparently, the COG is more time consuming than HOG, and one possible future work is to optimize the efficiency of COG. In our implementation, the current parameter settings are quite coarse, thus we presume that the recognition rate will grow if the distance is much larger or the angular partition is refined further.

## 8.4 Realistic movie recognition

Although the recognition rate of the Hollywood dataset is still low, yet the methods we compared give promising solutions for future research. *Dollár et al.*'s extraction method is good at collecting interest points around body parts that have significant motions, and the number of points is rich. *Laptev et al.*'s method is a multi-scale based approach and good at extracting points from sequences shot with different camera zooms, however, the number of detected points is easily affected by the volume of the sequences and the background noise. Currently there is no perfect way to conquer the problems occurred in realistic human action recognition, therefore, we think the future work is to focus on getting rid of the disturbance from irrelevant background activities.

## 8.5 Plan for publication

Based on the conclusions and recommendations we made above, our future work is to achieve descriptor fusion by making use of both structural distribution and appearance information. The efficiency and accuracy of approaches applied in Hollywood dataset also needs to be emphasized in our future work.

Currently, we are planning to submit the 'shape context' part to the "The 1st International Workshop on Video Event Categorization, Tagging and Retrieval". We also plan to submit the rest of the thesis to the Journal of "Computer Vision and Image Understanding".